\documentclass[a4paper,11pt]{article}
\pdfoutput=1
\usepackage{jcappub}[=2023-03-02]
\usepackage{ml-proofs}

\usepackage{microtype}

\usepackage{makecell}
\usepackage{collcell}
\newcolumntype{G}{>{\collectcell\gape}{c}<{\endcollectcell}}
\newcolumntype{A}{>{\collectcell\gape}{l}<{\endcollectcell}}
\newcolumntype{Z}{>{\collectcell\gape}{r}<{\endcollectcell}}

\usepackage[labelformat=simple]{subcaption}
\usepackage{caption}
\usepackage{lscape}

\title{Multiwavelength analysis of Galactic Supernova Remnants}

\author[a,1]{P.~Sharma,\note{Corresponding author.}}
\emailAdd{sharma@ijclab.in2p3.fr}

\author[b]{Z.~Ou,}

\author[a]{C.~Henry-Cadrot,}

\author[a]{C.~Dubos}

\author[a]{and~T.~Suomij\"arvi}
\emailAdd{tiina.suomijarvi@ijclab.in2p3.fr}

\affiliation[a]{Universit\'e Paris-Saclay, CNRS/IN2P3,\\
IJCLab, 91405 Orsay, France}
\affiliation[b]{School of Physics and Astronomy, Sun Yat-sen University,\\
519082 Zhuhai, China}

\abstract{The origin of Galactic Cosmic Rays~(CRs) and the possibility of Supernova Remnants (SNRs) being potential CR accelerators is still an open debate. The charged CRs can be detected indirectly by the $\gamma$-ray observatories through the $\pi^0$ production and consequent decay, leading to the generation of high-energy $\gamma$-rays. The goal of the study is to identify qualitative and quantitative trends in favour of hadronic scenario and search for SNRs which could be potential accelerators up to PeV energies (PeVatrons).

We have performed a Multiwavelength (MWL) study using different radiative models to evaluate the hadronic contribution. The spectral energy distributions~(SEDs) of selected SNRs are modeled using the Naima~\cite{Naima_zabalza} package. Two different radiative scenarios are considered, pure leptonic and lepto-hadronic scenarios and different methods are used to evaluate their importance.

This study shows that the lepto-hadronic scenario is favored for most SNRs. Two particular indicators of hadronic contribution come from the data around the $\pi^0$ production threshold and the data above a few TeV\@. The hard rise at the $\pi^0$ production threshold cannot be explained by leptonic processes. More data in this region would be valuable for these studies. For some SNRs, an important hadronic contribution is observed up to a few TeV, thus making them promising PeVatron candidates. In this high-energy region where the leptonic processes are expected to be suppressed, more data is required to help distinguish between the leptonic and hadronic origin of $\gamma$-ray emission. In the future, we intend to use the obtained model parameters to simulate data for CTA and assess its capability to identify~PeVatrons.}

\begin{document}
\maketitle\flushbottom

\section{Introduction}
For almost a century, Supernova Remnants (SNRs) have been considered to be the predominant sources of Galactic Cosmic Rays~(CR), which consist primarily of relativistic protons and ions~\cite{Baade_Zwicky_1934}. This is supported by the fact that only $\sim$ 10--20 \%  of the kinetic energy of SN explosion is sufficient to produce the observed CR spectrum. Well-established theoretical models explain how particles can be accelerated in SN shock waves through diffusive shocks~\cite{DSA_2012}, also called the first-order Fermi acceleration, up to the energies approaching $10^{15}$\,eV\@. Out of all the features of the CR spectrum~\cite{Lipari_2020}, the one which is the most relevant to this study is the knee at $10^{15}$\,eV, mainly attributed to protons and light elements. In addition to the prominent knee, evidence for a second knee has been found around $10^{16}$\,eV due to Iron or similar heavy elements~\cite{Second_knee_2011}. Astrophysical sources that can accelerate particles up to the knee are called PeVatrons. The energetic particles which are produced at the site of acceleration are diffused by the interstellar magnetic fields, making it impossible to point to the source directly. However, when these particles interact with the surrounding medium, high-energy $\gamma$-rays are produced, which allow us to trace back to the CR sources.

   The current instruments allow us to make observations spanning from radio to TeV bands. The non-thermal emission is mostly due to synchrotron, bremsstrahlung, and inverse Compton radiation from CR electrons. Additionally, secondary $\gamma$-rays due to $\pi^0$ production and the subsequent pion-decay ($p + p \rightarrow \pi^0 \rightarrow \gamma + \gamma $) provide indirect evidence for CR ions.
   Observations by \emph{Planck}\footnote{\url{https://sci.esa.int/web/planck}.} provide information on the microwave region of the spectra due to non-thermal electrons. Other instruments such as \emph{Chandra},\footnote{\url{https://chandra.harvard.edu/}.} \emph{Swift}\footnote{\url{https://www.swift.ac.uk/}.} and \emph{Integral}\footnote{\url{https://sci.esa.int/web/integral}.} (INTErnational Gamma-Ray Astrophysics Laboratory) have played a major role in scanning the sky in the X-ray domain. Instruments such as \emph{Fermi},\footnote{\url{https://fermi.gsfc.nasa.gov/}.} HAWC\footnote{\url{https://www.hawc-observatory.org/}.} (High-Altitude Water Cherenkov Observatory), AS$\gamma$,\footnote{\url{https://www.icrr.u-tokyo.ac.jp/em/index.html}.} and LHAASO\footnote{\url{https://directory.eoportal.org/web/eoportal/satellite-missions/l/lhaaso}.} (Large High Altitude Air Shower Observatory) further provide valuable data in the $\gamma$-ray domain. The imaging atmospheric Cherenkov telescopes~(IACT) such as  VERITAS\footnote{\url{https://veritas.sao.arizona.edu/}.} (Very Energetic Radiation Imaging Telescope Array System), H.E.S.S.\footnote{\url{https://www.mpi-hd.mpg.de/hfm/HESS/}.} (High Energy Stereoscopic System) and MAGIC\footnote{\url{https://astro.desy.de/gamma_astronomy/magic/index_eng.html}.} (Major Atmospheric Gamma Imaging Cherenkov Telescope) have provided a wealth of observations in the TeV range of the spectrum. H.E.S.S. has already shown the possibility of detecting galactic PeVatrons from which the energy spectra extend beyond 10\,TeV~\cite{Cristofari_2018}. The upcoming observatory CTA\footnote{\url{https://www.cta-observatory.org/}.} (Cherenkov Telescope Array) will further complement the high-energy observations by present instruments. Due to its higher sensitivity, accuracy, and wide energy coverage, we expect a significant increase in the quantity and quality of the observations in the very high-energy $\gamma$-ray domain.

   Gamma rays are produced in this domain by hadronic or leptonic interactions. Three main ways of disentangling between these components are 1) studying the $\gamma$-ray spectra well beyond $\sim$ 10 TeV, where one expects a decline of inverse Compton radiation due to the Klein-Nishina effect~\cite{longair_2011}; 2) studying older SNRs ($\geq$ 100 yrs) where electrons are expected to have exhausted their energy and hence cannot reach the multi-TeV energies~\cite{Aharonian_electron}; 3) studying SNRs which are known to be near high density ($>$ 10 cm$^{-3}$) molecular clouds, which increase the production of $\pi^0$ due to hadronic interactions~\cite{Slane_2014}.

   This study aims to evaluate the hadronic contribution of selected Galactic SNRs and search for potential CR accelerators. To achieve this, we performed an MWL study of 9 SNRs using leptonic and lepto-hadronic models. Section~\ref{sec:dat-ml-select} describes the selection criteria for choosing the SNRs and the source of data. The methodology and the tools used in the study are explained in section~\ref{TypSec_3}. Section~\ref{TypSec_4} provides a few exemplar results, which are discussed in detail. All the results can be found in the appendix. Three quantitative analyses used in the study are discussed in section~\ref{TypSec_5}. In section~\ref{TypSec_6}, we compare our results to those previously obtained for the selected SNRs. The findings of this paper are summarized in section~\ref{TypSec_7}.
\label{sec:intro}

\section{Data selection}
\label{sec:dat-ml-select}

In this study, we focused on two types of SNRs: shell-type and interacting SNRs. The shell-type SNRs are often observed as bright rings formed due to the radiation from the shell of shocked material. The interacting SNRs are the remnants of massive stars which explode and interact with the molecular clouds in which they were born~\cite{Slane_2014}.

   All the shell and interacting type SNRs listed in TeVCat~\cite{TeVCat} (29 sources) were chosen as a preliminary list of SNRs. From this list, the SNRs were further divided into different datasets: the silver and the golden datasets, depending on the availability of MWL observations. The silver dataset includes the sources which only have high-energy data. The golden dataset contains the final SNRs for which high-energy data is available in gamma-cat~\cite{gammapy_2019} and additional data found in SSDC,\footnote{\url{https://tools.ssdc.asi.it/SED/}.} Green catalog~\cite{Green_2019}, \emph{Chandra} X-ray catalog\footnote{\url{https://hea-www.harvard.edu/ChandraSNR/snrcat_gal.html}.} and other literature.  Finally, for each selected SNR, its flux and uncertainty were collected, and an MWL spectral energy distribution~(SED) was constructed.

 For choosing the observations, a few selection criteria were applied. First, the latest observations or surveys were favoured. Hence observations from the newest available catalogues were chosen. Second, we have eliminated observations from the UV or IR bands to avoid contamination of the source from nearby stars and dust. Third, we have only chosen instruments for which the Field of View (FoV) was comparable to or larger than the size of the chosen SNR.

After these considerations, we are left with the 9 SNRs listed in table~\ref{tab:TypeSources}. All the data points and their origin are listed in table~\ref{tab:OriginData} in the appendix. For each SNR, we provide information about its type (table~\ref{tab:AllData}), which can be found in TeVCat and age (table~\ref{tab:TypeSources}) along with the references.

\begin{table}
\renewcommand{\arraystretch}{1.1}    \centering
    \begin{tabular}{ |p{3cm}|p{1.5cm}|p{3cm}|p{3.4cm}| }
      \hline
      Source & Type &  Age (years) & Magnetic field ($\mu$G) \\
      \hline
      Cassiopeia A & Shell & 340~\cite{CasA_1999} & 400~\cite{helder2008characterizing}\\
      \hline
      CTB 37A & Shell & 6000~\cite{Fermi_ctb37a_2020} & 30~\cite{Fermi_ctb37a_2020}\\
      \hline
      HAWC J2227+610  & Int. & 4700--5700~\cite{Albert_2020} & 100~\cite{bao2021} \\
      \hline
      HESS J1731-347  & Shell & 2000--6000~\cite{Cui_2019} & 50~\cite{HessJ1731Shell}\\
      \hline
      IC 443 & Int. & 700--2300~\cite{bib2005ICRC_HESS} & 100~\cite{zhang2018nustar}\\
      \hline
      RX J1713.7-3946 & Shell & 1580--2100~\cite{Tsuji_2016} & 14~\cite{RXJdoubt}\\
      \hline
      Vela Junior & Shell & 700~\cite{bib2001AIPC_vela} & 7~\cite{abdalla2018deeper}\\
      \hline
      W 28 & Int. & 57500--92500~\cite{bib2005ICRC_HESS} & 79~\cite{zirakashvili2017snrs}\\
      \hline
      W 49B & Int. & 1000--4000~\cite{refId0} & 200~\cite{brogan2001very}\\
      \hline
    \end{tabular}
    \caption{\label{tab:TypeSources}List of the chosen SNRs along with their type (Int. refers to SNRs interacting with the surrounding medium) and~age.}
\end{table}

\section{Analysis and tools}
\label{TypSec_3}
\subsection{Radiative models}
\label{TypSec_3.1}
Data from various instruments covering a large part of the electromagnetic (EM) spectrum encompass the non-thermal radiation emitted at the interaction site between the expanding SNR and the surrounding medium, along with magnetic and radiation fields. A detailed description of these interactions can be found in~\cite{longair_2011}.

Synchrotron radiation results from the acceleration of electrons streaming through the magnetic field. At times, a moving charged particle may be deflected by another charged particle causing it to decelerate. The moving particle loses kinetic energy, which is converted into bremsstrahlung radiation. Relativistic electrons emit inverse Compton radiation by up-scattering the low-energy photons in the background radiation, mainly Cosmic Microwave Background~(CMB) and infrared (IR) radiations to $\gamma$-ray energies. These together form the leptonic scenario where the radiation is due to electrons and positrons.

Another process that also leads to $\gamma$-ray emission at high energy is the pion-decay. Pion production occurs when accelerated protons interact with protons of the surrounding medium. The proton-proton collisions produce $\pi^0$ mesons (p + p $ \rightarrow$ p + p + $\pi^0 $) which subsequently decay into two $\gamma$-rays ($\pi^0 \rightarrow \gamma + \gamma $) having the energy of $\sim$ 70\,MeV in the $\pi^0$ rest frame. The threshold energy (kinetic energy of proton colliding with another stationary proton) for $\pi^0$-production is 280\,MeV\@. Observing the $\pi^0$ bump in the photon spectrum provides evidence for high-energy protons accelerated by the source. This corresponds to the hadronic scenario.

\subsection{Leptonic and lepto-hadronic scenarios}
\label{TypSec_3.2}
Disentangling the different nature of $\gamma$-ray emission from leptons and hadrons continues to be an ongoing matter of interest. If emission was ascribed to be mostly hadronic compared to leptonic, it would mean that the source accelerates protons and nuclei primarily. However, studies indicate that given the normal interstellar medium conditions, the hadronic contribution would dominate only if electrons are less than approximately 0.1 $\%$ abundant compared to protons~\cite{Gabici_2016}. Hence, it is not straightforward to attribute very-high-energy $\gamma$-ray emission from SNRs to hadronic processes.

In our attempt to understand which scenario is supported by the available observations, we have fitted observed SEDs. The fit was done using two different models: a lepto-hadronic model, where the photons are assumed to be produced by a mixture of leptonic and hadronic processes, and a purely leptonic model, where photons are assumed to be produced only through leptonic processes.

In this study, we have considered protons for the hadronic processes and electrons for the leptonic processes. We assume the energy distribution of both protons, $f_p(E)$ and electrons, $f_e(E)$ to follow an exponential cut-off power-law (ECPL) with $\beta$ as the cut-off exponent. The distributions for protons and electrons are given below:
\begin{align} f_p(E) & = A_p \times \left(\frac{E}{E_0}\right)^{\alpha_p} \times \exp ({-E/E_{\mathrm{COP}})^ \beta} \label{eq:1},  \\[5pt]
 f_e(E) & = A_e \times \left(\frac{E}{E_0}\right)^{\alpha_e} \times \exp ({-E/E_{\mathrm{COE}})^ \beta } \label{eq:2},  \end{align}
with:
\begin{itemize}
    \item $A_p$: the amplitude of the proton distribution
    \item $A_e$: the amplitude of the electron distribution
    \item  $\alpha_{e}$: the electron spectral index
    \item $\alpha_{p}$: the proton spectral index
    \item $E_{\mathrm{COE}}$: the electron energy cut-off
    \item $E_{\mathrm{COP}}$: the proton energy cut-off
    \item $\beta$: the cut-off exponent
\end{itemize}

The reference energy has been set to $E_0 = 1$\,TeV\@. The parameter $K_{ep} = A_e/A_p$ describes the ratio of electrons with respect to protons at 1\,TeV\@. This allows us to redefine $A_p$ as $A$.

Since $\pi^0$ production depends on the density of protons in the surrounding medium $N_{H}$ and the synchrotron radiation depends on the magnetic field $B$ in the radiating region, we need to add two more parameters to obtain the overall fit. Finally, eight free parameters are required to theoretically build SEDs: $A, \alpha_{e}, \alpha_{p}, E_{\mathrm{COE}}, E_{\mathrm{COP}}, Kep, B, n_H$.

In contrast to the lepto-hadronic model, the leptonic model only considers bremsstrahlung, inverse Compton, and synchrotron processes. Defining the electron energy distribution as equation~\eqref{eq:2}, we find that only four free parameters are needed to theoretically build the SEDs: $A, \alpha_{e}, E_{\mathrm{COE}}, B$.

 Since electrons suffer radiative losses, one would expect the electron distribution to differ from a simple ECPL\@. To account for this behaviour, we have studied the effect of using an Exponential Cutoff Broken Power Law~(BPL) function for electron distribution for three SNRs; RX J1713.7-3946, IC 443, and W 49B\@. We re-fit these SNRs by using BPL with the parameters found in the literature~\cite{RXJdoubt,bibW49B_H_F,Ackermann_2013}. The values for the break energy used in the fit were 2\,TeV, 0.1\,TeV, and 0.2\,GeV for RX J1713.7-3946, W 49B, and IC 443, respectively. By using BPL for electron distribution, we got slightly better fit results for these SNRs.

 In order to estimate the break energy~\cite{RXJdoubt}, we used the equation~\eqref{eq:3}.
\begin{equation} E_b \simeq 1.25 \left(\frac{B}{100 \mu G} \right)^{-2} \left(\frac{t_0}{10^3 yr} \right)^{-1} \text{TeV} \label{eq:3}.  \end{equation}

From the equation, we can see that the break energy is dependent mainly on the magnetic field and the age of the SNR ($t_0$). The average values of SNR age have been taken from table~1, and the B$_{\mathrm{fit}}$ values used have been derived from Naima fitting (see appendix~\ref{TypSec_B}). For the three SNRs, we found the break energies to vary from 0.1 to 1\,TeV due to large uncertainty in their age. The calculated values are in fair agreement with the values of the break energy used for the fit. However, while it is clear that a break is expected for the studied SNRs due to the cooling process, the accuracy of the data used for our study is not sufficient for a detailed analysis of the shape of the electron spectrum.

 Additionally, we studied the effect on the fit by using a more general form for the electron distribution. This was done by varying the value of $\beta$ as 0.5, 1, and the effect of leaving it~free.

 Statistical analysis of the results of both studies showed that there was no significant qualitative and quantitative impact on the overall results. Therefore, we conclude that the addition of more parameters to the model is not justified. In the following, we have used ECPL with $\beta$ = 1 for both electron and proton distributions.

\subsection{The Naima fitting tool}
\label{TypSec_3.3}
We have used Naima~\cite{Naima_zabalza} to build theoretical SEDs. It is a Python package consisting of radiative models which allow us to compute the non-thermal emission from populations of relativistic electrons and protons and perform Monte Carlo Markov Chain~(MCMC)~\cite{MCMC_Naima} fitting to the observed spectra. All the models take as input the parent particle distribution function. Different sets of parameters, called priors, need to be initialized to sample the spectral model parameters' space. The priors encode our knowledge about the probability distribution of the model parameters. The run returns two sets of parameters: the best-fit parameters and their uncertainties and the median value of each parameter obtained at the end of the chain of the different initial sets. Naima gives the option to choose between two priors; normal and uniform priors. After studying the fit using both types of priors, we concluded that the uniform prior is a better option yielding a better fit to the observations. Supporting our decision is the fact that the uniform prior assigns equal probability to each point in the wide range of parameters, whereas the normal prior maximizes the probability of values around the mean represented by initial parameters, yielding lower probabilities for the parameters at the edge of the parameter range. Thus using normal priors would require us to have more precise knowledge about physical parameters in contrast to uniform priors, which only require a wide enough range to explore the parameter space effectively. Additionally, using normal priors increases the risk of getting stuck in a local minimum.

Naima assumes that the measurements and uncertainties provided for the SED building are correct, Gaussian, and independent.\footnote{\url{https://buildmedia.readthedocs.org/media/pdf/naima/saving/naima.pdf}.} The likelihood of a set of parameters is thus computed using the equation~\eqref{eq:4}:
\begin{equation} L = \prod_{i=1}^N \frac{1}{\sqrt{2 \pi \sigma_i^2}} \exp{\left(-\frac{(S(\Vec{p};E_i)-F_i)^2}{2 \sigma_i^2}\right)} \label{eq:4},  \end{equation}
with N being the total number of measurements, $F_i$ the measured flux at the energy $E_i$, and $\sigma_i$ the associated uncertainty. $S(\Vec{p};E_i)$ is the theoretical flux at a particular energy $E_i$ given the assumed spectral model and the set of parameters $\Vec{p}$.

Calculations can be significantly simplified by studying $\log(L)$ instead of $L$. This gives the equation~\eqref{eq:5}:
\begin{equation} \log(L) = \sum_{i=1}^N \left(-\frac{(S(\Vec{p};E_i)-F_i)^2}{2 \sigma_i^2}\right) \label{eq:5}.  \end{equation}

The $\ln\mathcal{L}$ can be related to $\chi^2$ parameter using the equation:
\begin{equation} \chi^2=-2\log{L} \label{eq:5-ml-a}.  \end{equation}

This makes the maximization of log-likelihood equivalent to the minimization of  $\chi^2$.

For this study, the synchrotron emission model has been used to fit the low-energy contribution of the observed SED\@. At high energy, either purely leptonic (bremsstrahlung \& inverse Compton) or a mixture of lepto-hadronic models (bremsstrahlung, inverse Compton \& pion-decay) have been considered. In most cases, it has been seen that the inverse Compton process is predominantly responsible for the TeV $\gamma$-ray emission~\cite{Albert_2020}.

For modeling the bremsstrahlung radiation, the default values of solar system abundance were provided by Naima~\cite{Naima_zabalza}.
Additionally, we have assumed that the inverse Compton radiation arises solely due to the interaction between energetic electrons and the photon seed field from the CMB and the Far InfraRed background (FIR)\@. Although other fields exist, FIR radiation due to dust emission in the Milky Way-type galaxies provides the most important contribution for $\gamma$-ray photons ranging from GeV to TeV~\cite{FIR_galaxy}. For the fields considered, the default values set by the Naima package have been used i.e., a temperature of 2.72 K, 26.5 K, and energy density of 0.261\,eVcm$^{-3}$ 0.415\,eVcm$^{-3}$ for the CMB and the FIR radiations,~respectively.

In order to compare the pure leptonic and lepto-hadronic models, we make use of the Bayesian Information Criterion~(BIC) to see which model is better at explaining the observed spectrum~\cite{BIC}. The BIC analysis has been used for model selection in various other papers before~\cite{Ambrogi_2019,Fang_2022,dePalma,Albert_2020}. The BIC quantity is defined using the equation~\eqref{eq:6}:
\begin{equation} BIC = k \log(n) - 2 \log(L_{\mathrm{best}}) \label{eq:6},  \end{equation}
with $k$ being the number of parameters in the model, $n$ the number of data points, and $L_{\mathrm{best}}$ the likelihood associated with the best-fit parameters for a given model. BIC has the advantage of being independent of the prior and measures the parameterized model's efficiency in predicting the data. Furthermore, during model fitting, the addition of parameters may increase the likelihood. The BIC analysis allows us to take this into account by introducing a penalty term for the number of parameters in the model. However, it is limited by the fact that a valid data sample much larger than the number of parameters in the model is required.

When comparing two models, the model with a lower BIC is preferred. The BIC value increases with an increasing value of k. Thus lower BIC implies either fewer explanatory variables, better fit, or both. We have decided to consider two different models, leptonic and lepto-hadronic, and the magnitude of the $\Delta$BIC =  $BIC_{\mathrm{leptonic}} - BIC_{\text{lepto-hadronic}} $  will allow us to determine which model is better. The resulting  $\Delta$BIC value is to be interpreted as evidence against pure leptonic being the better model. If the resulting $\Delta$BIC value is larger than 2, then the model with the lower BIC value is slightly preferred. If the $\Delta$BIC value is greater than 6, then the model with a lower BIC value is favoured. No firm conclusion can be drawn between the two values but it would tend toward favoring the model with a lower BIC value.

\subsection{Model constrains}
\label{TypSec_3.4}

Having short-listed the SNRs with sufficient data points, in order to start SED fitting, we first need to limit the range of possible values that parameters can take. This is done to avoid biased results and to ensure that the output parameters lie within the physically acceptable range.
Experimental results have shown that the spectral index of CRs lies between $-1.5$ $\leq$ $\alpha$ $\leq$ $-3$ both for electrons and protons. Therefore, we have chosen $\alpha$ to vary between these values.
The natural constrain on the $K_{ep}$ value also arises from the fact that it is easier for electrons to be absorbed compared to protons while traversing through medium. Hence we put a constraint of $0 < K_{ep} \leq 1$. In order to constrain B and $n_H$ values, we studied various articles\footnote{\cite{HessJ1731Shell,Ackermann_2013,Brandt_2013,Yamauchi_2014,bibW49B_H_F,Zeng_2017,zirakashvili2017snrs,Ahnen_2017,Guo_2018,Sushch_2018,Tanaka_2018,Zhang_2019,Xin_2019,Phan_2020,Abeysekara_2020,Fermi_ctb37a_2020,Kokusho_2020,Fujita_2021,Tang_2021,fukui2021pursuing}.} and decided to choose as an upper limit the value that was greater than all those found in the literature. Zero was chosen as a lower limit for both variables.
Finally, for the amplitude, which is a re-normalizing constant with no obvious constraints, the boundary was chosen to allow the fit to converge given all the other constraints.
The constraints applied to each parameter are listed in the table~\ref{tab:boundaryCondition}.

\begin{table}
\renewcommand{\arraystretch}{1.2}    \centering
    \begin{tabular}{ |p{7.5cm}|p{3cm}|p{3cm}| }
      \hline
      Parameter & Lower limit & Upper limit \\
      \hline
      log10(Amplitude) & 40 & 60 \\
      \hline
      Electron spectral index & 1.5 & 3 \\
      \hline
      Proton spectral index  & 1.5 & 3 \\
      \hline
      log10(electron energy cut-off in TeV) & $-5$ & 6 \\
      \hline
      log10(proton energy cut-off in TeV) & $-5$ & 6 \\
      \hline
      B ($\mu$G) & 0 & 300 \\
      \hline
      Kep & 0 & 1 \\
      \hline
      $n_H$ ($\text{cm}^{-3}$) & 0 & 1000 \\
      \hline
     \end{tabular}
    \caption{\label{tab:boundaryCondition}Parameter range used as input prior~information.}
\end{table}

\section{Results}
\label{TypSec_4}

\subsection{Examples of fit results using different scenarios}
\label{TypSec_4.1}
In this section, we discuss a few examples of fit results obtained using both the lepto-hadronic and leptonic scenarios. In all the SED plots, the black dots represent the observations (table~\ref{tab:OriginData} in appendix~\ref{TypSec_A}) from several instruments for which the value of the flux measurement at a given energy is known with a precise uncertainty. Red triangles are used to differentiate those observations for which the flux measurements are uncertain and hence have been plotted as an upper limit.

The contribution of the radiative emissions from the sources is shown using different curves, each representing a different radiative process. The solid yellow curve depicts the theoretical synchrotron flux; the green dot-dash curve represents the theoretical flux due to bremsstrahlung radiation; the orange curve with a dotted pattern gives the theoretical flux pertaining to inverse Compton, and the red dashed curve corresponds to emission due to pion-decay. The solid blue curve shows the combined contribution of all these emissions.

\begin{figure}
\centering
\begin{subfigure}{0.4\linewidth}
  \includegraphics[width=1.0\linewidth, height = 0.8\linewidth]{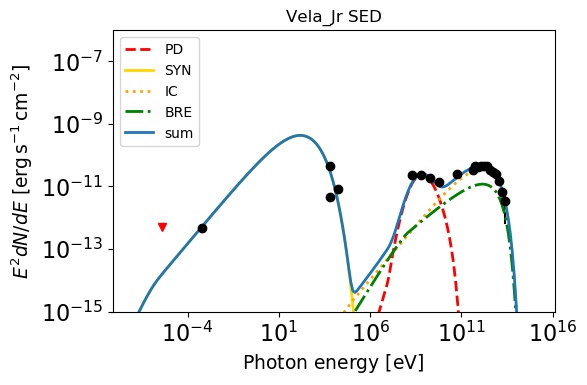}
\end{subfigure}
\begin{subfigure}{0.4\linewidth}
  \includegraphics[width=1.0\linewidth, height = 0.8\linewidth]{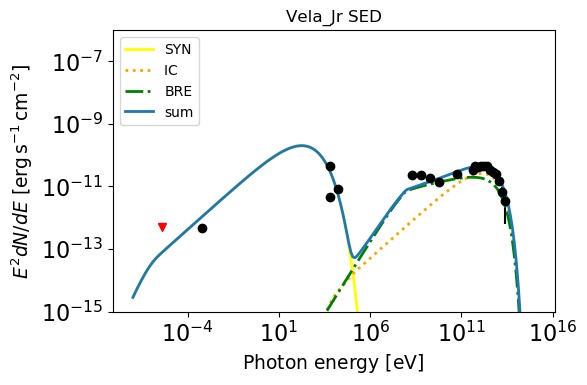}
\end{subfigure}
\caption{MWL SED of the SNR Vela Junior modeled by using the hadronic scenario (left) and the leptonic scenario (right). Abbreviations stand for PD: pion-decay radiation, SYN: Synchrotron radiation, IC: Inverse Compton, BRE:~Bremsstrahlung.\label{VelaH&L}}
\end{figure}

As a first example, the fit result for the SNR Vela Jr. assuming a lepto-hadronic and a leptonic scenario is shown in figure~\ref{VelaH&L}. Through this example, we can see that a fit with a pure leptonic model leaves some data points at the beginning of the $\gamma$-ray domain, thus giving a poor fit. Whereas, with the help of pion-decay radiation in the lepto-hadronic scenario, the increased emission in the GeV domain can be accounted for.


\begin{figure}
\centering
\begin{subfigure}{0.4\linewidth}
  \includegraphics[width=1.0\linewidth, height = 0.8\linewidth]{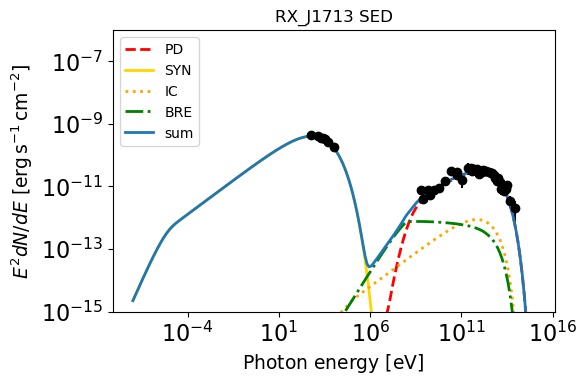}
\end{subfigure}
\begin{subfigure}{0.4\linewidth}
  \includegraphics[width=1.0\linewidth, height = 0.8\linewidth]{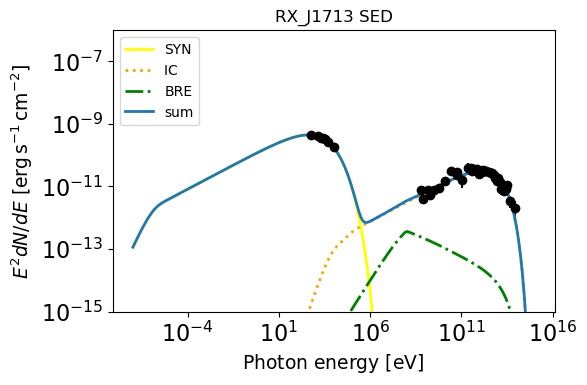}
\end{subfigure}
\caption{MWL SED of the SNR RX J1713.7-3946 modeled by using the hadronic scenario (left) and the leptonic scenario (right). For meaning of abbreviations see caption of figure~\ref{VelaH&L}.\label{RXJH&L}}
\end{figure}

The next example is the SNR RX J1713.7-3946. The result of the fit for the lepto-hadronic and the leptonic scenario is shown in figure~\ref{RXJH&L}. This example has been chosen because it contrasts sharply with the previous one. Looking at these figures, we can see that both lepto-hadronic and leptonic scenarios reproduce the observations well. This is further confirmed by the similarity of the obtained likelihood value (Max Log-Likelihood --- see appendix~\ref{TypSec_B}).

\subsection{Obtained fit parameters}
\label{TypSec_4.2}
 The fit results for all the studied SNRs together with the parameters obtained, assuming the two different scenarios, can be found in the appendix~\ref{TypSec_B} (figures~\ref{HAWC_H}--\ref{W49B_L} and tables~\ref{tab:ParaFitHawcJ2227}--\ref{tab:ParaFitW49B}). Table~\ref{tab:Pararange} gives the output range for each parameter obtained by fitting the lepto-hadronic or leptonic models to the observed spectra. A similar parameter range can be observed when comparing the two scenarios. However, for a lepto-hadronic scenario, higher magnetic fields are needed to match the observed flux compared to leptonic scenarios, which always predict a lower magnetic field irrespective of its ability to fit data well. Overall, the parameters lie well within the acceptable values recorded in various papers.

Assuming the lepto-hadronic scenario, the cut-off value of electron (COE) and proton (COP) distributions for different SNRs have been tabulated in table~\ref{tab:naimacutoff}. The maximum cut-off value for protons is $\sim$ 447\,TeV for the SNR HAWC J2227+610.

\begin{table}
\renewcommand{\arraystretch}{1.2}    \centering
    \begin{tabular}{ |p{3cm}|p{3cm}|p{3cm}|  }
        \hline
        Parameter & Lepto-hadronic & Leptonic \\
        \hline
        $e_{\mathrm{alpha}}$ & 1.61 -- 2.80 & 1.34 -- 2.66 \\
        \hline
        $p_{\mathrm{alpha}}$ & 1.64 -- 2.49 & Nil \\
        \hline
        $E_{{\mathrm{cut}},e}(\text{TeV})$ & 0.69 -- 14.80 & 0.17 -- 630.96\\
        \hline
        $E_{{\mathrm{cut}},p}(\text{TeV})$ & 0.02 -- 446.68 &  Nil\\
        \hline
        B($\mu$G) & 14.12 -- 960.61 & 0.29 -- 86.08 \\
        \hline
        $K_{ep}$ & 0.00160 -- 0.024 & Nil\\
        \hline
        $N_{H}(\text{cm}^{-3})$ & 1.70 -- 163.21 & 0.07 -- 205.86  \\
        \hline
    \end{tabular}
        \caption{\label{tab:Pararange}Range of fit parameters obtained for all studied SNRs using Naima for different radiative~scenarios.}
\end{table}

\begin{figure}
    \centering
    \includegraphics[width = 0.8\linewidth, height=7.4cm]{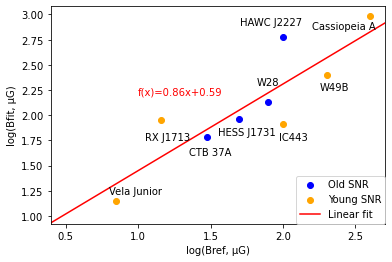}
    \caption{The value of the magnetic field obtained from the fit as a function of  the reference magnetic field. The age and the magnetic field of the SNRs have been taken from table~1, and the references therein. The old SNRs are indicated by blue circles, while the young SNRs are represented by orange~circles.\label{mag_comp}}
\end{figure}

\enlargethispage*{\baselineskip}\relax
Figure~\ref{mag_comp} shows the magnetic field reported in the literature for various sources as a function of the magnetic field obtained from the fit. A rather good correlation between these two values is observed. However, in the present study, the scarcity of data around the synchrotron peak limits the accurate determination of the magnetic field. On the other hand, the fit at the highest energies is rather well constrained with existing data which allows us to determine the contribution of inverse Compton and indirectly constrain the electron distribution and therefore also estimate the B-field values.
\pagebreak

\begin{figure}
    \centering
    \includegraphics[width = 0.8\linewidth, height=7.4cm]{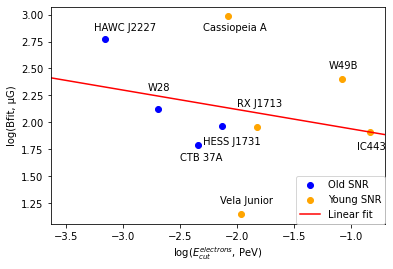}
    \caption{The magnetic field as a function of the electron energy cut-off value obtained from the fit. The age of the SNRs has been taken from table~1, and references therein. The old SNRs are indicated by blue circles, while the young SNRs are represented by orange~circles.\label{age_elec}}
\end{figure}

The radiative losses of electrons should depend on the age of the SNR, the younger the SNR, the shorter the cooling time for electrons. In figure~\ref{age_elec}, one can observe that the young SNRs have higher cut-off energies compared to old SNRs. This could indicate either a shorter cooling time or a more powerful acceleration for young SNRs.

\begin{figure}
    \centering
    \includegraphics[width = 0.8\linewidth, height=7.4cm]{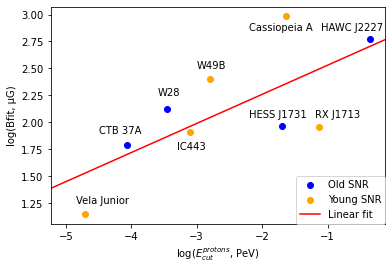}
    \caption{The magnetic field as a function of the cut-off energy of protons obtained from the fit. The age of the SNRs has been taken from table~1, and references therein. The old SNRs are indicated by blue circles, while the young SNRs are represented by orange~circles.\label{age_pro}}
\end{figure}

We have also studied the correlation between the magnetic field and the cut-off energy of protons, as shown in figure~\ref{age_pro}. We can observe that the proton energy cut-off value increases with an increasing magnetic field. On the top right corner of the plot, we can see the SNRs RX J1713.7-3946, Cassiopeia A, HESS J1731-347, and HAWC J2227+610, which are identified as potential PeVatron candidates in our study, as will be discussed later. Additionally, contrary to electrons, we don't see any trend as a function of the age of the SNRs.

Finally, we saw no correlation between the number density and the type of SNR\@.  Similarly, no correlation was observed between the magnetic field and the type of SNR nor between the proton energy cut-off and the type of SNR.

The experiments such as CTA, HAWC, and LHAASO would allow us to observe much higher energy photons and give more constrain to sign hadron acceleration up to higher energies. The parameters obtained in this study for the radiative models can be used to simulate data for these instruments to study their detection capability for observing PeVatron~sources.

\begin{table}
\renewcommand{\arraystretch}{1.3}    \centering
    \begin{tabular}{ |p{3.5cm}|p{3.5cm}|p{3.5cm}| }
      \hline
      SNR & $E_{\mathrm{COE}}$ (TeV) & $E_{\mathrm{COP}}$ (TeV)\\
      \hline
      HAWC J2227+610 & $0.69\substack{+0.004 \\ -0.004}$ & $446.68\substack{+0.07 \\ -0.07}$ \\
      \hline
      Cassiopeia A & $8.32\substack{+0.004 \\ -0.004}$ & $23.40\substack{+0.01 \\ -0.01}$ \\
      \hline
      CTB 37A & $4.50\substack{+0.02 \\ -0.02}$ & $0.08\substack{+0.05 \\ -0.05}$ \\
      \hline
      IC 443 & $1.46\substack{+0.04 \\ -0.03}$ & $0.80\substack{+0.04 \\ -0.06}$ \\
      \hline
      RX J1713.7-3946 & $14.80\substack{+0.04 \\ -0.03}$ & $74.10\substack{+0.04 \\ -0.05}$ \\
      \hline
      HESS J1731-347 & $7.41\substack{+0.04 \\ -0.03}$ & $20.40\substack{+0.04 \\ -0.06}$ \\
      \hline
      Vela jr & $10.80\substack{+0.02 \\ -0.02}$ & $0.02\substack{+0.08 \\ -0.05}$ \\
      \hline
      W 28 & $2.02\substack{+0.01 \\ -0.01}$ & $0.35\substack{+0.01 \\ -0.02}$  \\
      \hline
      W 49B & $3.10\substack{+0.01 \\ -0.02}$ & $1.60\substack{+0.00 \\ -0.01}$ \\
      \hline
    \end{tabular}
    \caption{\label{tab:naimacutoff}Electron ($E_{\mathrm{COE}}$) and proton ($E_{\mathrm{COP}}$) distribution cut-off energies obtained from the fit of the studied~sources.}
\end{table}

\section{Quantitative analysis}
\label{TypSec_5}
Three different methods were used to quantify the hadronic contribution: likelihood comparison, BIC criterion, and contribution as a function of energy bins.
\subsection{Likelihood comparison}
\label{TypSec_5.1}
A method that can be used to distinguish which model better describes the observations is the likelihood comparison. Naima returns the likelihood of the best-fit parameters as an output of the fitting process. Likelihoods were obtained for fits using the two different models, pure leptonic and lepto-hadronic models.

We have compared the likelihood of each leptonic and lepto-hadronic model with the sum of the likelihood of both models, given by r. The equations used are shown by eq.~(\ref{eq:7-ml-a-ml-}) and eq.~(\ref{eq:7-ml-b-ml-}).
\begin{subequations}
  \begin{align}r_{\text{lepto-hadronic}} & =  1 - \frac{\chi^2_{L+H}}{\chi^2_{L}+\chi^2_{L+H}} \label{eq:7-ml-a-ml-}.  \\[5pt]
    r_{\mathrm{leptonic}} & = 1 - \frac{\chi^2_{L}}{\chi^2_{L}+\chi^2_{L+H}} \label{eq:7-ml-b-ml-}.  \end{align}
\end{subequations}
\begin{figure}
    \centering
    \includegraphics[width = 0.65\linewidth, height=12.1cm]{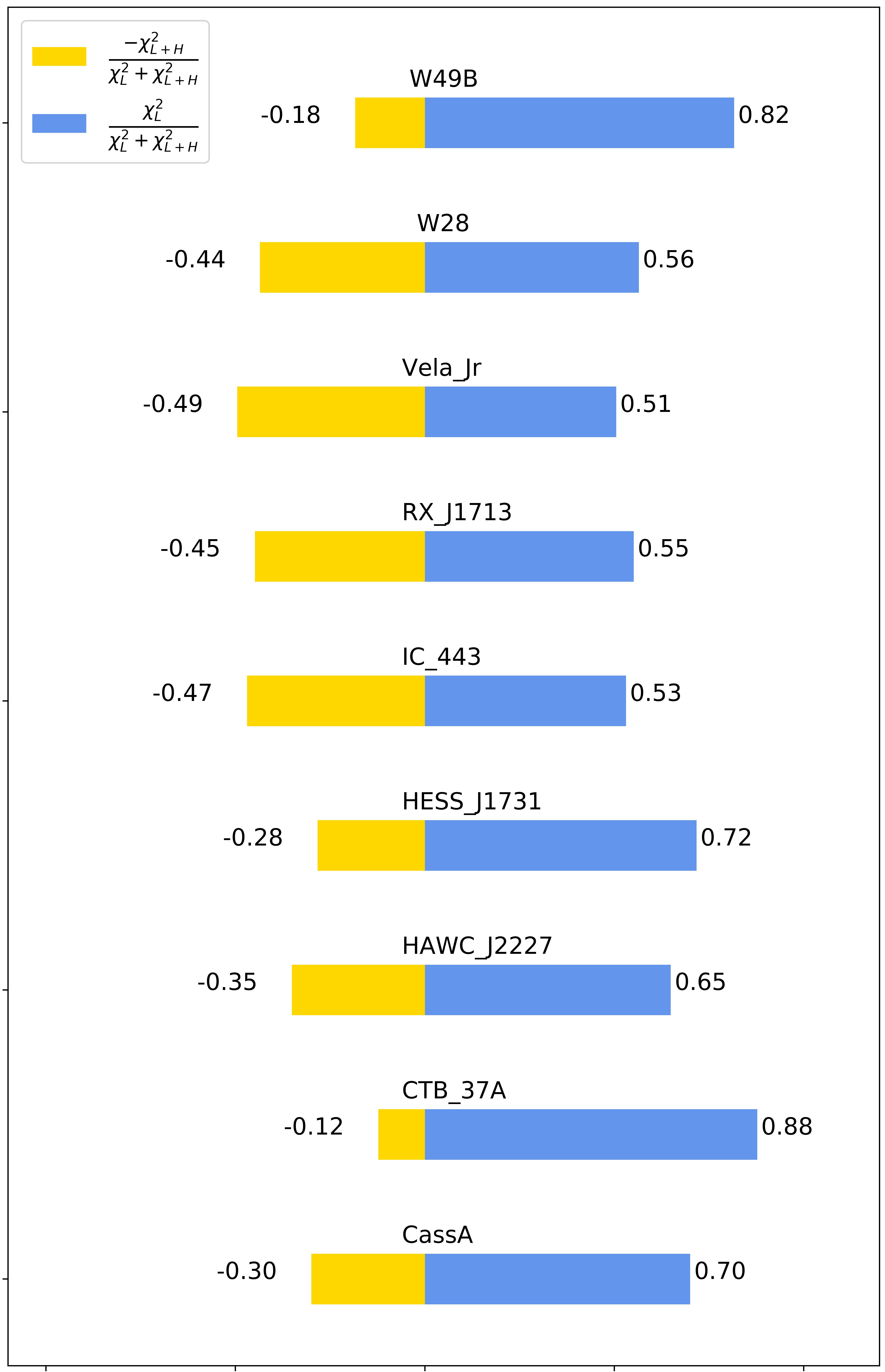}
    \caption{Statistical comparison between leptonic and lepto-hadronic models using likelihood~method.\label{ComparisonModel}}
\end{figure}

The results of this comparison can be found in figure~\ref{ComparisonModel}. In the figure for each source fitted with a particular model, the blue bar corresponds to eq.~(\ref{eq:7-ml-a-ml-}), and the yellow bar corresponds to eq.~(\ref{eq:7-ml-b-ml-}). The total length of the bar is constrained to 1. The bar tends to $+1$ if the likelihood of the lepto-hadronic model is greater than that of the leptonic model and to $-1$ otherwise.

Comparing the two scenarios also shows more confidence in the lepto-hadronic scenario than in the purely leptonic scenario for each source since the ratio $r_{Lepto-hadronic}$ is greater than 0.5 for all sources. The most uncertain ones are Vela-Junior and IC 443. RX J1713.7-3946 and HAWC J2227+610, although uncertain, tend more towards the lepto-hadronic scenario.
However, unlike the previously discussed qualitative analysis, this method clearly allows us to conclude that for W 49B, HESS J1731-347, CTB 37A, and Cassiopeia A, the lepto-hadronic model is~favoured.

\subsection{BIC criterion}
\label{TypSec_5.2}
From the BIC value of the leptonic and lepto-hadronic scenarios obtained for each source, we can compute $\Delta$BIC value and apply the previously introduced criterion to determine which scenario is favored for each source. The $\Delta$BIC for different sources are shown in the table~\ref{tab:BIC-ml-per-ml-sources}.

\begin{table}
    \centering
\renewcommand{\arraystretch}{1.2}
    \begin{tabular}{ |p{3cm}|c|c|c| }
      \hline
      Source & Leptonic & Lepto-hadronic & $\Delta$ BIC\\
      \hline
      Cassiopeia A & 6851 & 2956 & 3895 \\
      \hline
      CTB 37A & 194 & 47 & 147 \\
      \hline
      HAWC J2227+610 & 118 & 82 & 36 \\
      \hline
      HESS J1731-347 & 52 & 41 & 11 \\
      \hline
      IC 443 & 235 & 219 & 16 \\
      \hline
      RX J1713.7-3946 & 112 & 106 & 6 \\
      \hline
      Vela Junior & 5088 & 4994 & 94 \\
      \hline
      W 28 & 207 & 172 & 35 \\
      \hline
      W 49B & 1158 & 285 & 873 \\
      \hline
    \end{tabular}
    \caption{\label{tab:BIC-ml-per-ml-sources}BIC values for considered sources. $\Delta BIC$ is defined by BIC(Leptonic) -- BIC(Lepto-hadronic).}
\end{table}

Recalling that for $ \Delta$BIC $>2$ the model with the smaller BIC is slightly favored, and for $\Delta$BIC $> 6$ the model with the smaller BIC is clearly favored, we can see that lepto-hadronic scenario is clearly favored for the majority of the sources given our initial assumptions for radiative models.

However, for the SNR RX J1713.7-3946, we get a $\Delta$BIC value of 6. Although it appears to favour the lepto-hadronic scenario, it is difficult to obtain a firm conclusion.

\subsection{Hadronic contribution as a function of energy}\label{RadSec}

\begin{figure}
\centering
\includegraphics[width=1\textwidth]{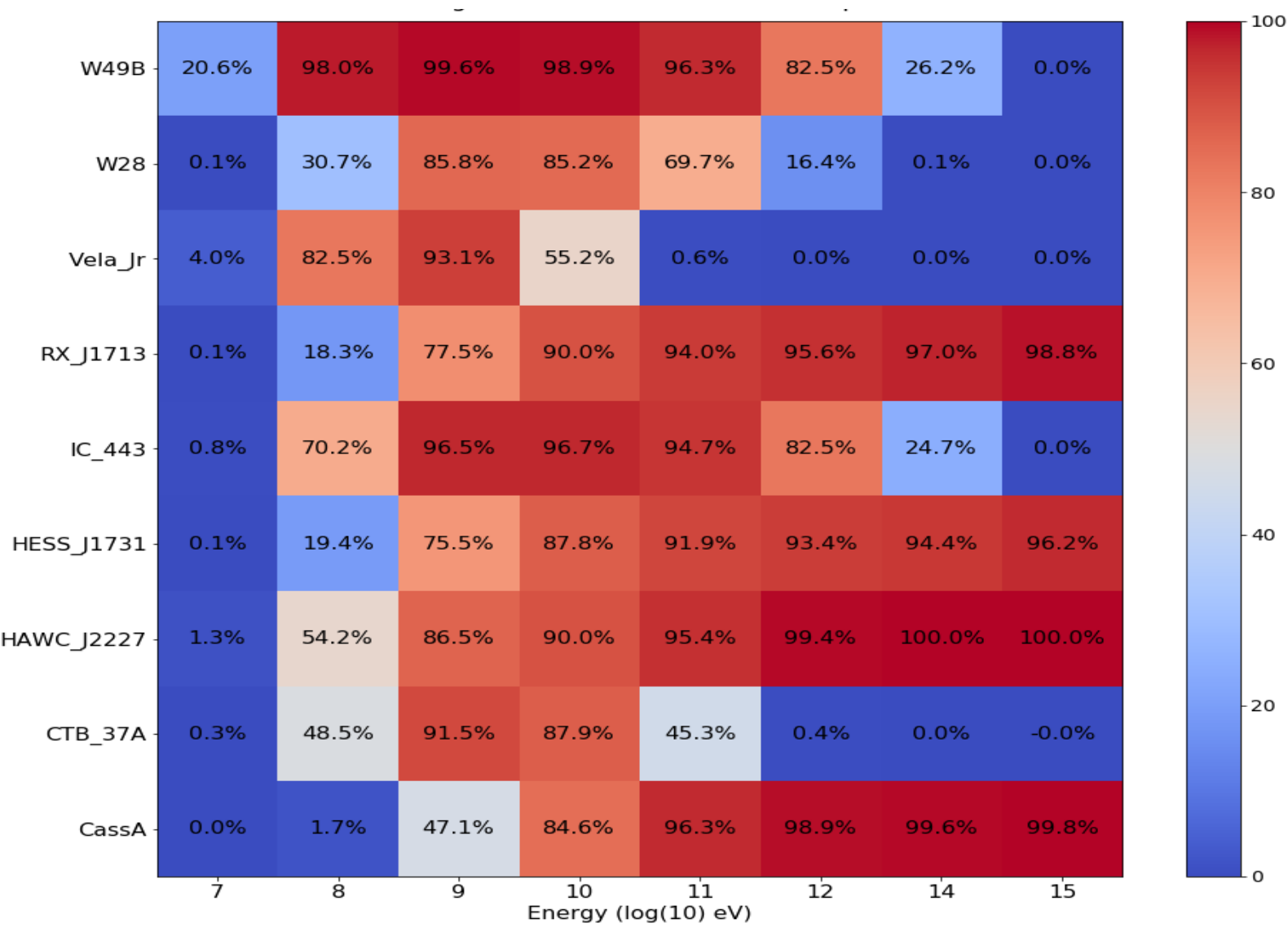}
    \caption{Flux integrated over one order of magnitude energy bins. The number depicts the percentage of the hadronic component in each energy~bin.\label{Flux_comp_bins}}
\end{figure}

In the previous studies, only the overall shape of the SED resulting from various radiative contributions was taken into account. However, a hadronic contribution can become more important when focusing on specific energy ranges. Hence using a lepto-hadronic scenario for each SNR, an integration over the flux ranging from $10^{7}$ to $10^{15}$\,eV was performed.
This method allows us to compare the relative importance of the leptonic and hadronic processes at each order of magnitude. Figure~\ref{Flux_comp_bins} shows the results of this binned integration for each~source.

This plot shows that in the range of $E \in [10^{9} - 10^{11}]$\,eV, the hadronic process dominates for every source. Additionally, most sources show a contribution from the leptons up to the highest energies (TeV range). An interesting result that can be highlighted from this plot is that the SNRs RX J1713.7-3946, HESS J1731-347, HAWC J2227+610, and Cassiopeia A show fit results dominated by hadron contribution up to the highest energies. This opens up a promising possibility for these four sources to be PeVatrons.

\section{Discussion}
\label{TypSec_6}

\enlargethispage*{\baselineskip}\relax

\subsection{Selected SNRs}
\label{TypSec_6.1}

This section discusses the results obtained for each SNR and compares them to the previous studies. It must be pointed out that exact comparison is often difficult for several reasons. In various publications, authors invoked their own fitting models, each with its own assumptions, to reproduce the observed flux. Additionally, in each case, different datasets have been used, thus making direct comparisons difficult.

In the following discussion, the information on the type and age of SNRs has been referred from papers shown in table~\ref{tab:TypeSources}.
\begin{itemize}
    \item Cassiopeia A\\
    Also known as G111.7-2.1, this young shell-type SNR has been previously studied by the MAGIC collaboration (\cite{Ahnen_2017} and references therein). It has been observed to expand supersonically, producing a shock wave where its high magnetic field can be amplified, leading to the possibility of CR acceleration up to PeV energies. A year-scale variability in the X-ray filaments suggests a magnetic field in the order of mG\@. The authors combined \emph{Fermi}-LAT and MAGIC data and concluded that $\gamma$-ray emission from 60\,MeV to 10\,TeV can be attributed to the population of high-energy protons with an energy cut-off at 10\,TeV.

    The $\Delta$BIC of $\sim 3895$ obtained from our MWL fits clearly favors the lepto-hadronic scenario. In our case, the proton energy cut-off was found to be $\sim$ 20\,TeV, which agrees with the paper referred to above. Furthermore, the fit result using the lepto-hadronic model gives us a magnetic field of $\sim$ 960 $\mu$G.

    \item CTB 37A\\
    Also referred to as G348.5+0.1, this source was detected by the \emph{Fermi} collaboration~\cite{Brandt_2013}. This complex source is interacting with several dense molecular clouds (\cite{Fermi_ctb37a_2020} and references therein). Previous studies conclude that the MWL observation is mostly due to the re-acceleration of preexisting cosmic rays by radiative shocks in the adjacent clouds.

    In our study, the $\Delta$BIC value of $\sim 147$ clearly favors the lepto-hadronic scenario. Hadronic contribution is most prominent around the GeV energy domain. In addition to this, in our study, additional data points from HESS indicate leptonic contribution above $10 $\,TeV (figure~\ref{Flux_comp_bins}). CTA will provide a better angular resolution to resolve the complex source into individual components.

\looseness=-1
    \item HAWC J$2227+610$\\
    Also known as G106.3+2.7, this source is a cometary medium-aged SNR with a tail in the Southwest and a head containing the PSR J2229+6114 in the Northeast~\cite{bao2021}. In a recent paper by the HAWC collaboration~\cite{Albert_2020}, $\gamma$-rays above 100\,TeV were reported. Despite combining results from VERITAS and HAWC, the authors could not confirm the origin of high-energy photons. The instrument AS$\gamma$ has also reported observation of high-energy gamma rays up to 100\,TeV~\cite{yang2021}. A MWL fit with data from the instrument Tibet AS+MD~\cite{nature_2021}, shows a spectral index of $1.76\substack{+0.02 \\ -0.03}$ and an exponential cut-off at $446.68\substack{+0.07 \\ -0.07}$\,TeV\@. These parameters are similar to the parameters found in this study. Additionally, the authors correlate the $\gamma$-ray emission above 10\,TeV with a MC which further supports the hadronic scenario. Furthermore, a recent paper by LHAASO collaboration~\cite{bib2021Natur} confirms HAWC J2227+610 to be a PeVatron. In our study, the $\Delta$BIC obtained was $\sim 40$ which favors the lepto-hadronic scenario. All these facts lead towards making HAWC J2227+610 the most promising PeVatron candidate amongst all the SNRs in this study.

    \item HESS J1731-347\\
    Also called G353.6-0.7, this source is a shell-type medium-aged SNR\@. It has been studied by the H.E.S.S. collaboration using MWL emission detected in radio, X-ray, and $\gamma$-rays~\cite{HessJ1731Shell}. The authors conclude that a large fraction of the kinetic energy of the explosion is transferred to accelerated protons. In a more recent paper,~\cite{Cui_2019}, the observed high-energy domain of the MWL SED indicates a hadronic contribution that can be explained using a shock-cloud collision model.

 In our study, the $\Delta$BIC of $\sim 11$ also supports the lepto-hadronic scenario with a strong contribution from hadrons in the high-energy domain extending up to $10^{14}$\,eV.

    \item IC 443\\
     Also known as SNR 189.1+03.0, this source is a shell-type medium-aged SNR\@. The characteristic pion-decay feature in the $\gamma$-ray spectra of the two SNRs, IC 443 and W44, has been reported by the \emph{Fermi} collaboration~\cite{Ackermann_2013} providing evidence that CR protons are accelerated in these SNRs. In both sources, the spectra exhibit a break at 200\,MeV that is related to the $\pi^0$ threshold.

 For our study, only IC 443 was considered due to insufficient data available for W44. The MWL study yields the $\Delta$BIC of $\sim 16$, which clearly favors the lepto-hadronic scenario. The SED shows a hadronic contribution dominating around $10^9$ to $10^{11}$\,eV.

    \item RX J1713.7-3946\\
 Among all the shell-type SNRs, the young SNR RX J1713.7-3946 is one of the brightest TeV sources~\cite{RXJdoubt}. The H.E.S.S. collaboration performed an MWL study, but neither hadronic, leptonic nor a mixture of both could explain the data unambiguously. However, the paper of ref.~\cite{Gabici_2014} suggests that the $\gamma$-ray emission from RX J$1713.7$-$3946$ could be explained by the $\pi^0$ decay produced in hadronic CR interactions with a dense, clumpy gas embedded in the SNR shell.

 Our study reveals the $\Delta$BIC of $\sim 6$, which mildly favors the lepto-hadronic model. The prospect of observing RX J1713.7-3946 by CTA was studied in ref.~\cite{bib2017_RXJ}. The study predicts that measurements at high energy by CTA would yield further constraints on leptonic or hadronic scenarios using spectral and morphological studies.

    \item Vela Junior\\
    In our study, we have included \emph{Fermi} LAT data points starting from $2$ x $10^8$\,eV using SSDC data archive. Their inclusion leads to a pronounced hadronic bump around $10^9$\,eV, similar to that of HAWC J2227+610. The $\Delta$BIC being $\sim 94$ favors the lepto-hadronic scenario in our studies.

    \item W 28\\
    W 28, also known as SNR G$006.4$-$00.1$, is a mixed morphology old SNR\@. In its vicinity, various sources are presently yielding high-energy TeV gammas~\cite{Aharonian_2008}. In the paper, the authors conclude that the association of W 28 with nearby molecular clouds could yield high-energy $\gamma$-rays of hadronic origin. This SNR was also studied in ref.~\cite{Hanabata_2014}. The authors suggest that the interaction between the nearby molecular clouds and CRs escaping the SNR could explain emission from the source.

    In our study, we have used the high-energy data available in gamma-cat corresponding to HESS J1801-233. The \emph{Fermi} LAT counterpart taken from the SSDC catalogue coincides with the coordinates of the same source. We obtain the $\Delta$BIC of $\sim 35$, clearly favoring the lepto-hadronic scenario. The SED shows a strong hadronic component around $10^9$\,eV corresponding to the $\pi^0$ threshold.

    \item W 49B\\
    Also known as G043.3-00.2, this source is an old SNR with mixed morphology observed to be interacting with molecular clouds. The \emph{Fermi}-LAT collaboration jointly with the H.E.S.S. collaboration~\cite{bibW49B_H_F} found the spectral features to be indicative of the hadronic origin of the $\gamma$-ray emission.

    Our study gives the $\Delta$BIC value of $\sim 873$, strongly favoring the lepto-hadronic scenario. The best fit is obtained when the pion-decay component is present at around $10^9$\,eV.

\end{itemize}
\subsection{General features}
\label{TypSec_6.2}
In the following, we discuss some general features that can be seen in our results.

 All the SNRs were fit using a pure leptonic and a lepto-hadronic model. In some cases, SNRs such as RX J1713.7-3946, HESS J1731-347, and IC 443 can be satisfactorily fit using both models. However, in general, for a majority of the SNRs, a pure leptonic model fails to match the observations in both the GeV and TeV domains. Therefore, to generalize, all the SNRs can be fit using the lepto-hadronic model.
All the SNRs modeled using the lepto-hadronic scenario show a pion-decay bump.  However, this component is observed to peak at different energies. This led us to divide SNRs into two groups.

The first group contains SNRs in which the component peaks at $10^9$\,eV\@. This group includes W 28, Vela Jr, IC 443, CTB 37A, and W49 B\@. Clear proof of hadronic contribution can be found if data around the pion production threshold is available. Inverse Compton processes cannot explain the hard rise of the photon flux. Therefore more data around $\pi^0$ threshold energies would be valuable to provide a signature of hadronic contribution.

The second group contains the SNRs where the component peaks at $10^{12}$\,eV\@. This group comprises Cassiopeia A, HESS J1731-347, HAWC J2227+610, and RX J1713.7-3946. This important hadronic contribution at high-energy may be attributed to the harder slope of the parent proton distribution or the presence of heavier elements.

In the SNRs CTB 37A, RX J1713.7-3946, W49 B, and W 28, the model predicts a softer spectrum above $10^{13}$\,eV, which clearly deviates from the observations at high energies. It may also be a signature of the acceleration of heavier elements, which, although fewer in number, can contribute significantly at higher energies. However, more data would be required at energies beyond $10^{14}$\,eV to confirm these observations.

\subsection{PeVatron candidates}
\label{TypSec_6.3}

Referring to figure~\ref{Flux_comp_bins}, where the high-energy contributions at different bins are shown, we can see that Cassiopeia A, HESS J1731-347, HAWC J2227+610, and RX J1713.7-3946 have important hadronic contributions at the highest energies. This could hint toward the possibility of these SNRs being PeVatron sources.

\begin{figure}
    \centering
    \includegraphics[width = 0.8\linewidth,  height=7.7cm]{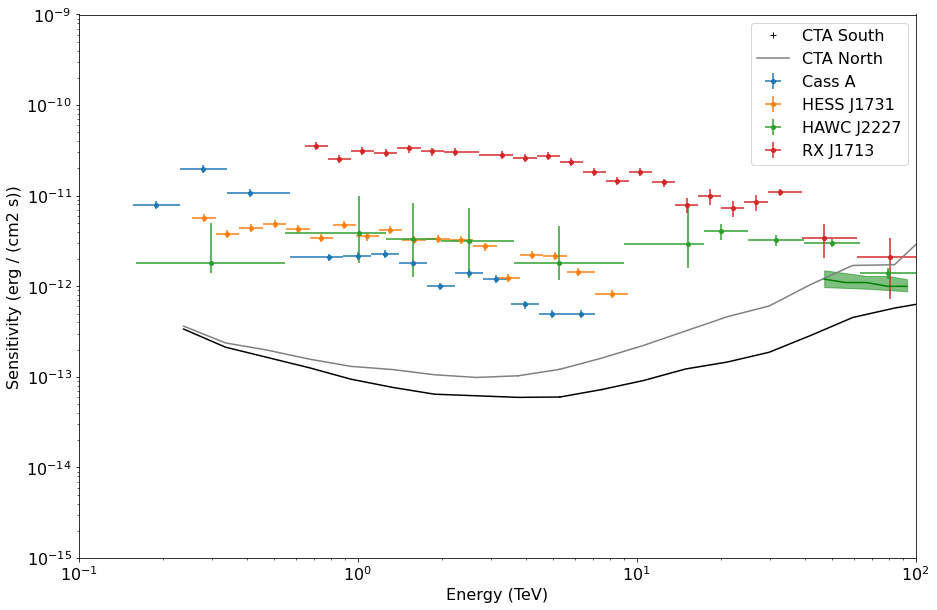}
    \caption{Sensitivity for CTA North (grey points) and South (black points) for a point source. The flux of four SNRs RX J1713.7-3946, HESS J1731-347, HAWC J2227+610 and Cassiopeia A have been compared to the CTA flux~sensitivity.\label{sensitivity}}
\end{figure}

Figure~\ref{sensitivity} shows the sensitivity curves of CTA North and CTA South together with the fluxes of the four SNRs mentioned above. The CTA sensitivity curves have been calculated by using CTAO Instrument Response Function~(IRF) --- prod5 version v0.1~\cite{cta_irf_2021}.\footnote{For the latest publically available IRF please visit --- \url{https://www.cta-observatory.org/science/cta-performance/}.} The telescopes have been assumed to be pointing parallel to each other. The performance has been estimated for a zenith angle of 20 degrees North and South with an observation time of 50h. The so-called Alpha configuration has been used for array layout, which includes 14 MSTs and 37 SSTs in the Southern site and 4 LSTs and 9 MSTs in the Northern site. The figure shows that CTA will clearly be able to add data to these four sources at high energy, allowing us to confirm emission processes at those energies. Additionally, CTA will provide flux with much less uncertainty compared to the present instruments and hence will help to constrain the spectral characteristics of the SNRs better. Simulations will be performed for CTA by using the model parameters obtained for these sources.

\section{Conclusions}
\label{TypSec_7}
In this study, we performed an MWL analysis of 9 galactic SNRs using publicly available data from recent experiments. To build MWL SEDs, we assumed that the observed SED was due to the synchrotron, bremsstrahlung, inverse Compton, and pion-decay processes. A lepto-hadronic scenario and a pure leptonic scenario were considered to model the observed flux. The Naima package was used to fit the observations. The fit results were analysed using three different methods: likelihood comparison, BIC criterion, and the estimation of the hadronic contribution as a function of energy bins. Furthermore, the results for each source were discussed in light of previous studies reported in various publications.

The Likelihood method used to compare the two scenarios showed more confidence in the lepto-hadronic scenario than in the purely leptonic scenario for most sources. Further proof of inclination towards the lepto-hadronic scenario was provided by determining the $\Delta$BIC value. However, solely based on $\Delta$BIC values, no firm conclusions can be drawn for the SNR RX J1713.7-3946.

The results obtained for binned energy contribution show that none of the sources have a proton distribution with cut-off energy $ \geq 10^{15}$\,eV, and therefore none can be classified as PeVatrons. However, four sources (SNR Cassiopeia A, RX J1713.7-3946, HAWC J2227+610, and HESS J1731-347) show important hadronic contributions at high energies. This opens up a promising possibility for these sources to be PeVatrons. In the future, we intend to use the range of output parameters obtained from Naima to get a set of physical input parameters to simulate synthetic spectra of SNRs at high energies to study the detection capabilities of CTA\@. In addition to these sources, the SNRs RX J1713.7-3946, CTB 37A, W 49B, and W 28 seem to show higher flux than predicted by the model above $10^{13}$\,eV when continuing to higher energies making them interesting candidates to study the acceleration of heavy elements. In the future, more data at high energy from instruments such as CTA, HAWC, and LHAASO would allow us to constrain hadronic contribution in this region better.

Most SNRs modeled using the lepto-hadronic scenario have a pion-decay bump. Here the hard rise of the photon flux cannot be explained by leptonic processes. Therefore more data around $\pi^0$ threshold energies would be valuable to provide a signature of hadronic contribution.

With an optimal sensitivity between $10^{11}$ and $10^{13}$\,eV and acute angular resolution in this energy range, CTA will be able to bring further clarity about whether or not a particular SNR is a PeVatron~\cite{Science_With_CTA}. In the future, we intend to consider the parent distribution, which includes heavier particles. Including heavier particles would lead to particles accelerated to energies higher than those of protons~\cite{refId0} corresponding to the so-called second knee.


\appendix
\section{Table for all data and their origin}
\label{TypSec_A}
Tables~\ref{tab:AllData} and~\ref{tab:OriginData} gives information on all the SNRs present in the preliminary list and on the origin of data points for the selected SNRs.
\begin{landscape}

\begin{table}
 \small
    \centering
                \begin{tabular}{|p{3cm}|p{1cm}|p{2.4cm}|p{2.5cm}|p{1.4cm}|p{1.3cm}|p{1.4cm}|p{1.3cm}|p{0.9cm}|p{1.1cm}|}
    \hline
    Source & Type & RA(hh mm ss) & DEC(dd mm ss)& Gal Long (deg) & Gal Lat (deg) & Distance (kpc) &Spectral Index & Silver Data & Golden Data \\
    \hline
    RX J1713.7-3946 & Shell &17 13 33.6	& -39 45 36 & 347.34 & -0.47 & 1 & 2.2 & Y & Y\\
    HESS J1731-347 & Shell & 17 32 03 &	-34 45 18 &	353.54 & -0.67 & 3.2 & 2.32 & Y&Y \\
    Tycho & Shell & 00 25 21.6 & +64 07 48	& 120.09 & 	1.4 & 	3.5	& 	1.95 & 	Y & N \\
    RX J0852.0-4622 & Shell & 08 52 00	& -46 22 00	& 266.28 & -1.24 & 0.2 & 1.81 & Y & Y\\
    IC 443 & Shell & 06 16 51 &	+22 30 11 & 189.07	& 2.92	& 1.5 & 3 & Y & Y \\
    RCW 86	& Shell & 14 43 02.16 & -62 26 56 & 315.44 & -2.32 & 2.5 & 2.54 & Y & N\\
    SN 1006 SW & Shell & 15 02 03.2	& -41 07 05 & 327.86 & 15.35 &	2.2 & 2.29 & N & N \\
    SN 1006 & Shell & 15 02 50 & -41 56 30 & 327.57 & 14.56	& 2.18	&	& Y & N\\
    SN 1006 NE & Shell & 15 04 03.4	& -41 48 11 & 327.84 & 14.56 & 2.2 & 2.35 &	N &	N \\
    HESS J1534-571 & Shell & 15 34 00 & -57 06 00 & 323.65 & -0.92 & & 2.51 & N & N \\
    HESS J1614-518 & Shell & 16 14 19.2 & -51 49 12 & 331.52 & -0.58 & 1.5 & 2.46 & N & N\\
    CTB 37B & Shell & 17 13 57.6 & -38 12 00 & 348.65 &0.38 & 13.2 & 2.65 & Y & Y \\
    HESS J1912+101 & Shell & 19 12 49 & +10 09 06 &44.39 & -0.07 & & 2.7 & Y & N \\
    0FGL J1954.4+2838 & Shell &	19 54 27.47	& +28 38 54.6 &	65.3 & 0.38	& 9.2 & & N & N \\
    SNR G106.3+02.7 & Shell & 22 27 59 & +60 52 37 & 106.35 & 2.71 & 0.8 & 2.29 & Y & Y \\
    SNR G106.3+02.7 & Shell & 22 27 59 & + 60 52 37	& 106.35 & 2.71 & 0.8 & 2.29 & Y & N\\
    Cassiopeia A & Shell & 23 23 13.8 & +58 48 26 &	111.71 & -2.13 & 3.4 & 2.3 & Y & Y\\
    LMC N132D & MC	& 05 25 02.20 & -69 38 39.0 & 280.31 & -32.78 & 50 & 2.4 & Y & N \\
    SNR G318.2+00.1 & MC & 14 57 46	 & -59 28 00 & 318.36 & -0.43 & 3.5	& 2.52 & N & N \\
    CTB 37A & MC & 17 14 15 & -38 31 27 & 348.42 & 0.14	& 7.9 &	2.3 & Y	& Y \\
    W 28 & MC & 18 01 42.2 & -23 20 06.0 & 6.66	 & -0.27 & 2 & 2.66 & Y & Y \\
    W 49B & MC & 19 11 07.3 & +09 09 37.0 & 43.32 &	-0.16 &	11.3 & 3.14 & Y & Y \\
    SNR G349.7+00.2 & MC & 17 17 57.8 & -37 26 39.6 & 349.72 & 0.17 & 11.5 & 2.8 & Y & N \\
    HESS J1745-303 & MC	& 17 45 02.10 & -30 22 14.00 & 358.71 & -0.64 & & 1.82 & Y & N \\
    HESS J1800-240A & MC & 18 01 57.8 & -23 57 43.2 & 6.14 & -0.63 & 2	& 2.55 & N & N \\
    HESS J1800-240B & MC & 18 00 26.4 & -24 02 20.4 & 5.9 & -0.37 & 2 & 2.5 & Y & N \\
    HESS J1800-240C & MC & 17 58 51.6 & -24 03 07.2 & 5.71 & -0.06 & 2 & 2.31 &	N &	N \\
    W 51 &  MC & 19 22 55.2	& + 14 11 27.6 & 49.12 & -0.36 & 5.4 & & N & N \\
    \hline
        \end{tabular}
        \caption{\label{tab:AllData}Information on the considered~SNRs.}
\end{table}

\begin{table}
    \centering
\renewcommand{\arraystretch}{1.2}

    \begin{tabular}{ |p{2.5cm}|p{1.5cm}|p{1.9cm}|p{1.5cm}|p{1.5cm}|p{1.5cm}|p{1.5cm}|p{1.5cm}|p{1.5cm}|  }
      \hline
      Source & SSDC\footnote{\url{https://tools.ssdc.asi.it/SED/}.} & Chandra\footnote{\url{https://hea-www.harvard.edu/ChandraSNR/snrcat_gal.html}.} & \cite{Green_2019} & \cite{gammapy_2019} & \cite{Abeysekara_2020} & \cite{VelaPoint} & \cite{Federici15} & \cite{Tanaka_2008}\\
      \hline
      Cassiopeia A & X & X & X & X & X & \_ & \_ & \_\\

      CTB 37A & X & X & X & X & \_ & \_ & \_ & \_\\

      HAWC J2227+610 & X & X & X & X & \_ & \_ & \_ & \_\\

      HESS J1731-347 & X & \_ & X & X & \_ & \_ & \_ & \_\\

      IC 443 & X & X & X & X & \_ & \_ & \_ & \_\\

      RX J1713.7-3946 & X & \_ & \_ & X & \_ & \_ & X & X\\

      Vela Junior & X & X & X & X & \_ & X & \_ & \_\\

      W 28 & X & X & X & X & \_ & \_ & \_ & \_\\

      W 49B & X & X & X & \_  & \_ & \_ & \_ & \_\\
      \hline
    \end{tabular}
    \caption{\label{tab:OriginData}Origin of the data points for the selected~SNRs.}
\end{table}
\end{landscape}

\section{Results for all SNRs}
\label{TypSec_B}
In the following figures~\ref{HAWC_H}--\ref{W49B_L}, we show the results of the fits obtained using either a lepto-hadronic model or a pure leptonic model for each source. Color codes show the different radiative processes. The solid yellow curve shows synchrotron emission (SYN), inverse Compton~(IC) radiation is given by the orange dotted curve, bremsstrahlung (BRE) radiation is represented by the green dot-dashed curve, and the red dashed curve depicts the pion-decay~(PD)\@. Lastly, the sum of all the radiations is given by the solid blue curve. Along with each SNR, a table~\ref{tab:ParaFitHawcJ2227}--\ref{tab:ParaFitW49B} is provided which gives the obtained parameters for both the lepto-hadronic and leptonic models.
\vspace*{\fill}
\pagebreak

\begin{figure}
    \centering
    \begin{subfigure}{1\linewidth}
        \centering
        \includegraphics[width=0.6\linewidth, height = 0.4\linewidth]{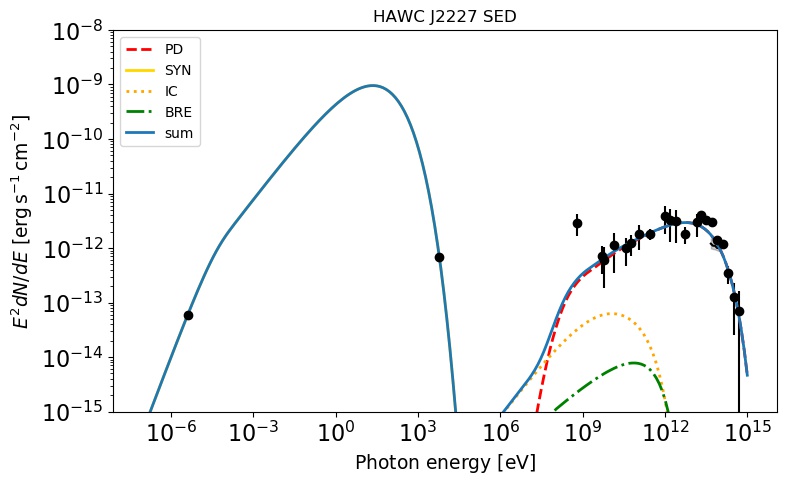}
        \caption{HAWC J2227+610 fitted by lepto-hadronic~model.\label{HAWC_H}}
    \end{subfigure}
    \begin{subfigure}{1\linewidth}
        \centering
        \includegraphics[width=0.6\linewidth, height = 0.4\linewidth]{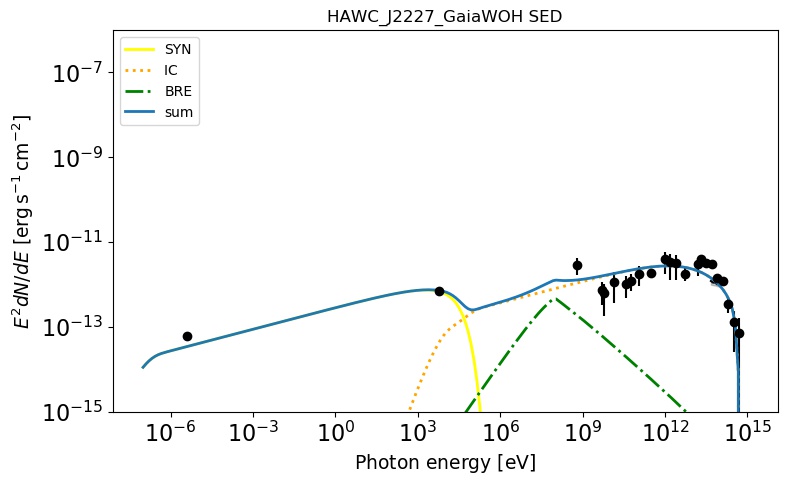}
        \caption{HAWC J2227+610 fitted by leptonic~model.\label{HAWC_L}}
    \end{subfigure}
    \caption{Naima fit results using different radiative scenarios for SNR HAWC J2227+610.}
\end{figure}

\begin{table}
\renewcommand{\arraystretch}{1.3}    \centering
    \begin{tabular}{ |p{6cm}|p{3cm}|p{3cm}|  }
      \hline
      Parameter & Lepto-hadronic & Leptonic \\
      \hline
      log10(Amplitude) & $47.02\substack{+0.001 \\ -0.001}$ & $45.43\substack{+0.03 \\ -0.04}$ \\
      \hline
      Electron spectral index & $1.69\substack{+0.02 \\ -0.03}$ & $2.66\substack{+0.03 \\ -0.05}$ \\
      \hline
      Proton spectral index  & $1.76\substack{+0.02 \\ -0.03}$ & X \\
      \hline
      electron energy cut-off (TeV) & $0.69\substack{+0.004 \\ -0.004}$ & $630.96\substack{+0.14 \\ -0.18}$ \\
      \hline
      proton energy cut-off (TeV) & $446.68\substack{+0.07 \\ -0.07}$ & X \\
      \hline
      B ($\mu$G) & $596.94\substack{+12.97 \\ -13.12}$ & $1.56\substack{+0.07 \\ -0.09}$ \\
      \hline
      Kep & $0.0047\substack{+0.0002    \\ -0.0004}$ & X \\
      \hline
      $n_H$ ($\text{cm}^{-3}$) & $1.70\substack{+0.09 \\ -0.10}$ & $0.08\substack{+0.04 \\ -0.03}$  \\
      \hline
      MaxLogLikelihood & $-27.45$ & $-50.74$ \\
      \hline
    \end{tabular}
    \caption{\label{tab:ParaFitHawcJ2227}Parameters obtained from the fit for HAWC J2227+610 using~Naima.}
\end{table}
\pagebreak

\begin{figure}
    \centering
    \begin{subfigure}{1\linewidth}
        \centering
        \includegraphics[width=0.6\linewidth, height = 0.4\linewidth]{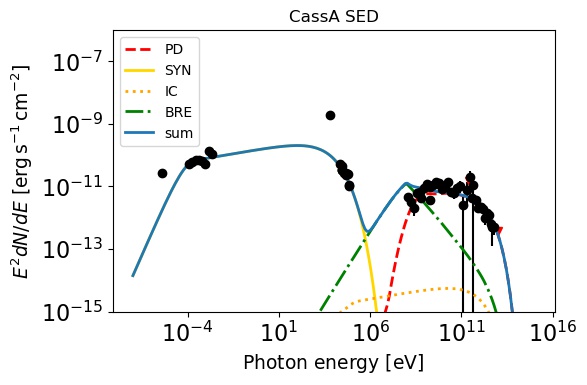}
        \caption{Cassiopeia A fitted by using lepto-hadronic~model.\label{Cass_H}}
    \end{subfigure}
    \begin{subfigure}{1\linewidth}
        \centering
        \includegraphics[width=0.6\linewidth, height = 0.4\linewidth]{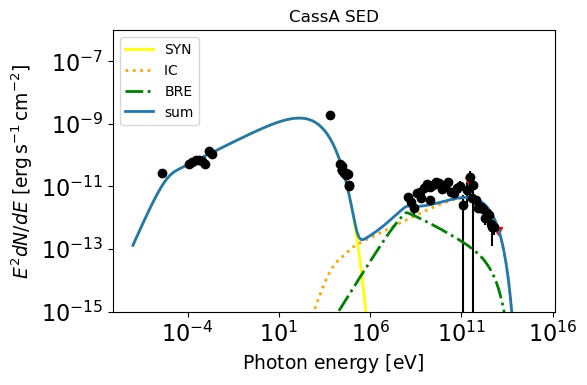}
        \caption{Cassiopeia A fitted by using leptonic~model.\label{Cass_L}}
    \end{subfigure}
     \caption{Naima fit results using different radiative scenarios for SNR Cassiopeia~A.}
\end{figure}

\begin{table}
\renewcommand{\arraystretch}{1.3}
\centering
    \begin{tabular}{ |p{6cm}|p{3cm}|p{3cm}|  }
      \hline
      Parameter & Lepto-hadronic & Leptonic \\
      \hline
      log10(Amplitude) & $47.03\substack{+0.0002 \\ -0.0004}$ & $46.90\substack{+0.002 \\ -0.003}$ \\
      \hline
      Electron spectral index & $2.80\substack{+0.001 \\ -0.002}$ & $2.42\substack{+0.02 \\ -0.03}$ \\
      \hline
      Proton spectral index  & $2.10\substack{+0.01 \\ -0.02}$ & X \\
      \hline
      electron energy cut-off (TeV) & $8.32\substack{+0.004 \\ -0.004}$ & $10.20\substack{+0.04 \\ -0.03}$ \\
      \hline
      proton energy cut-off (TeV) & $23.40\substack{+0.01 \\ -0.01}$ & X \\
      \hline
      B ($\mu$G) & $960.61\substack{+60.60 \\ -38.10}$ & $86.08\substack{+6.64 \\ -4.10}$ \\
      \hline
      Kep & $0.00160\substack{+0.00002 \\ -0.00001}$ & X \\
      \hline
      $n_H (\text{cm}^{-3}$) & $163.21\substack{+1.43 \\ -2.37}$ & $0.82\substack{+0.39 \\ -0.25}$ \\
      \hline
      MaxLogLikelihood & $-1461.56$ & $-3415.33$ \\
      \hline
    \end{tabular}
    \caption{\label{tab:ParaFitCassA}Parameters obtained from the fit for Cassiopeia A using~Naima.}
\end{table}
\pagebreak

\begin{figure}
    \centering
    \begin{subfigure}{1\linewidth}
        \centering
        \includegraphics[width=0.6\linewidth, height = 0.4\linewidth]{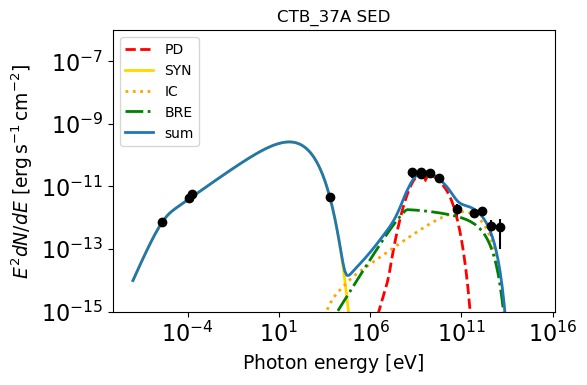}
        \caption{CTB 37A fitted by using lepto-hadronic~model.\label{CTB_H}}
    \end{subfigure}
    \begin{subfigure}{1\linewidth}
        \centering
        \includegraphics[width=0.6\linewidth, height = 0.4\linewidth]{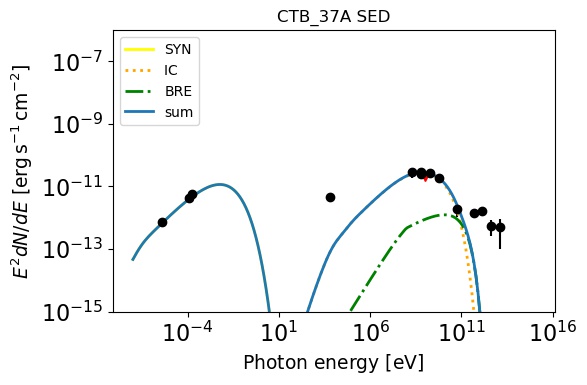}
        \caption{CTB 37A fitted by using leptonic~model.\label{CTB_L}}
    \end{subfigure}
    \caption{Naima fit results using different radiative scenarios for SNR CTB 37A.}
\end{figure}

\begin{table}
    \centering
\renewcommand{\arraystretch}{1.3}
    \begin{tabular}{ |p{6cm}|p{3cm}|p{3cm}| }
      \hline
      Parameter & Lepto-hadronic & Leptonic \\
      \hline
      log10(Amplitude) & $49.81\substack{+0.04 \\ -0.05}$ & $50.04\substack{+0.18 \\ -0.17}$ \\
      \hline
      Electron spectral index & $2.18\substack{+0.03 \\ -0.02}$ & $1.81\substack{+0.15 \\ -0.12}$ \\
      \hline
      Proton spectral index  & $2.00\substack{+0.04 \\ -0.04}$ & X \\
      \hline
      electron energy cut-off (TeV) & $4.50\substack{+0.02 \\ -0.02}$ & $0.17\substack{+0.07 \\ -0.07}$ \\
      \hline
      proton energy cut-off (TeV) & $0.08\substack{+0.05 \\ -0.05}$ & X \\
      \hline
      B ($\mu$G) & $61.19\substack{+2.36 \\ -3.06}$ & $3.20\substack{+0.59 \\ -0.59}$ \\
      \hline
      Kep & $0.0036\substack{+0.0002 \\ -0.0002}$ & X \\
      \hline
      $n_H$ ($\text{cm}^{-3}$) & $15.74\substack{+0.77 \\ -1.09}$ & $0.15\substack{+0.09 \\ -0.07}$ \\
      \hline
      MaxLogLikelihood & $-12.59$ & $-89.98$ \\
      \hline
    \end{tabular}
    \caption{\label{tab:ParaFitCTB37A}Parameters obtained from the fit for CTB 37A using~Naima.}
\end{table}
\pagebreak

\begin{figure}
    \centering
    \begin{subfigure}{1\linewidth}
        \centering
        \includegraphics[width=0.6\linewidth, height = 0.4\linewidth]{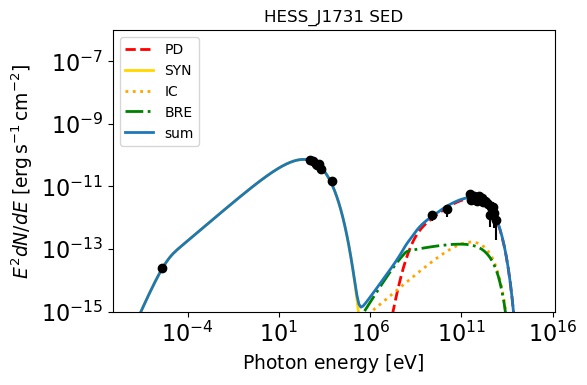}        \caption{HESS J1731-347 fitted by using lepto-hadronic~model.\label{J1731_H}}
    \end{subfigure}
    \begin{subfigure}{1\linewidth}
        \centering
        \includegraphics[width=0.6\linewidth, height = 0.4\linewidth]{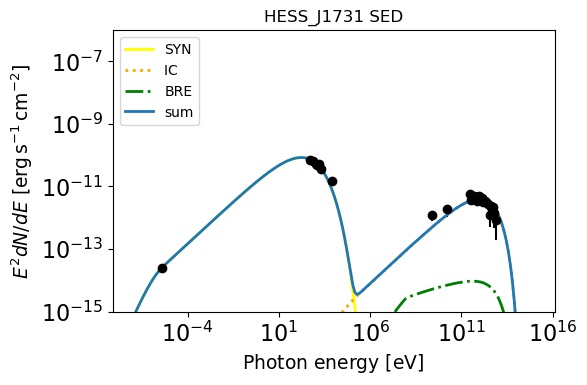}
        \caption{HESS J1731-347 fitted by using leptonic~model.\label{J1731_L}}
    \end{subfigure}
    \caption{Naima fit results using different radiative scenarios for SNR HESS J1731-347.}
\end{figure}

\begin{table}
    \centering
\renewcommand{\arraystretch}{1.3}
    \begin{tabular}{ |p{6cm}|p{3cm}|p{3cm}|  }
      \hline
      Parameter & Lepto-hadronic & Leptonic \\
      \hline
      log10(Amplitude) & $47.28\substack{+0.007 \\ -0.009}$ & $46.41\substack{+0.04 \\ -0.03}$ \\
      \hline
      Electron spectral index & $2.02\substack{+0.02 \\ -0.03}$ & $1.92\substack{+0.02 \\ -0.03}$ \\
      \hline
      Proton spectral index  & $1.64\substack{+0.04 \\ -0.04}$ & X \\
      \hline
      electron energy cut-off (TeV) & $7.41\substack{+0.04 \\ -0.03}$ & $13.10\substack{+0.05 \\ -0.05}$ \\
      \hline
      proton energy cut-off (TeV) & $20.40\substack{+0.04 \\ -0.06}$ & X \\
      \hline
      B ($\mu$G) & $91.98\substack{+3.63 \\ -4.37}$ & $18.86\substack{+1.07 \\ -1.03}$ \\
      \hline
      Kep & $0.0117\substack{+0.0005 \\ -0.005}$ & X \\
      \hline
      $n_H$ ($\text{cm}^{-3}$) & $45.16\substack{+2.04 \\ -2.16}$ & $0.23\substack{+0.27 \\ -0.15}$ \\
      \hline
      MaxLogLikelihood & $-7.01$ & $-17.76$ \\
      \hline
      \end{tabular}
    \caption{\label{tab:ParaFitHess1731}Parameters obtained from the fit for HESS J1731-347 using~Naima.}
\end{table}
\pagebreak

\begin{figure}
    \centering
    \begin{subfigure}{1\linewidth}
        \centering
        \includegraphics[width=0.6\linewidth, height = 0.4\linewidth]{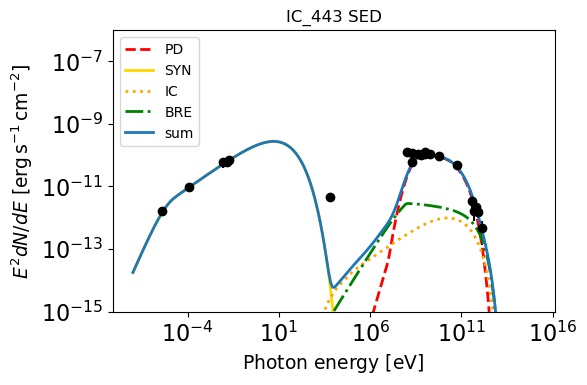}
        \caption{IC 443 fitted by using lepto-hadronic~model.\label{IC443_H}}
    \end{subfigure}
    \begin{subfigure}{1\linewidth}
        \centering
        \includegraphics[width=0.6\linewidth, height = 0.4\linewidth]{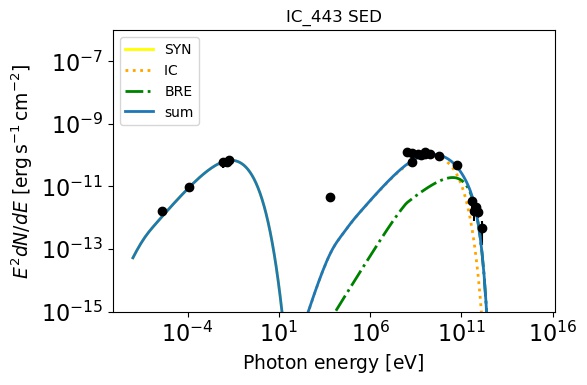}
        \caption{IC 443 fitted by using leptonic~model.\label{IC443}}
    \end{subfigure}
    \caption{Naima fit results using different radiative scenarios for SNR IC 443.}
\end{figure}

\begin{table}
    \centering
\renewcommand{\arraystretch}{1.3}
    \begin{tabular}{ |p{6cm}|p{3cm}|p{3cm}| }
      \hline
      Parameter & Lepto-hadronic & Leptonic \\
      \hline
      log10(Amplitude) & $48.51\substack{+0.03 \\ -0.03}$ & $49.01\substack{+0.1 \\ -0.1}$ \\
      \hline
      Electron spectral index & $2.18\substack{+0.01 \\ -0.01}$ & $1.63\substack{+0.05 \\ -0.06}$ \\
      \hline
      Proton spectral index  & $2.16\substack{+0.01 \\ -0.02}$ & X \\
      \hline
      electron energy cut-off (TeV) & $1.46\substack{+0.004 \\ -0.003}$ & $0.26\substack{+0.04 \\ -0.05}$ \\
      \hline
      proton energy cut-off (TeV) & $0.800\substack{+0.001 \\ -0.003}$ & X \\
      \hline
      B ($\mu$G) & $81.56\substack{+2.26 \\ -3.56}$ & $4.13\substack{+0.05 \\ 0.08}$ \\
      \hline
      Kep & $0.0037\substack{+0.0001 \\ -0.00007}$ & X \\
      \hline
      $n_H$ ($\text{cm}^{-3}$) & $16.75\substack{+0.35 \\ -0.53}$ & $1.23\substack{+0.26 \\ -0.21}$ \\
      \hline
      MaxLogLikelihood & $-97.10$ &$-109.87$ \\
      \hline
      \end{tabular}
    \caption{\label{tab:ParaFitIC443}Parameters obtained from the fit for IC 443 using~Naima.}
\end{table}
\pagebreak

\begin{figure}
    \centering
    \begin{subfigure}{1\linewidth}
        \centering
        \includegraphics[width=0.6\linewidth, height = 0.4\linewidth]{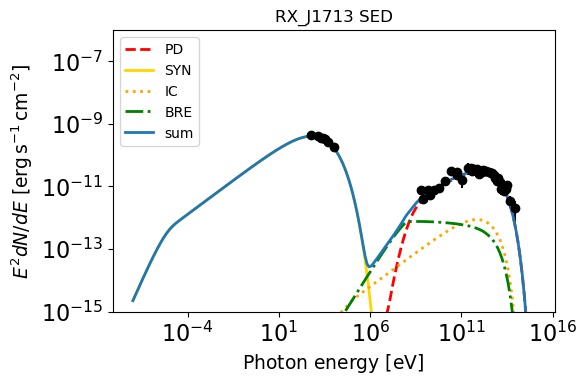}
        \caption{RX J1713.7-3946 fitted by using lepto-hadronic~model.\label{J1713_H}}
    \end{subfigure}
    \begin{subfigure}{1\linewidth}
        \centering
        \includegraphics[width=0.6\linewidth, height = 0.4\linewidth]{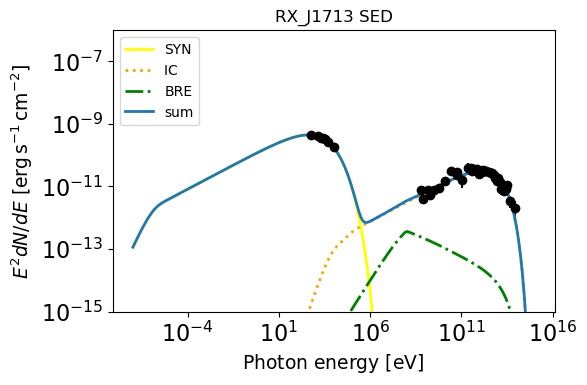}
        \caption{RX J1713.7-3946 fitted by using leptonic~model.\label{J1713_L}}
    \end{subfigure}
    \caption{Naima fit results using different radiative scenarios for SNR RX J1713.7-3946.}
\end{figure}

\begin{table}
    \centering
\renewcommand{\arraystretch}{1.3}
    \begin{tabular}{ |p{6cm}|p{3cm}|p{3cm}|  }
      \hline
      Parameter & Lepto-hadronic & Leptonic \\
      \hline
      log10(Amplitude) & $47.060\substack{+0.001 \\ -0.001}$ & $46.50\substack{+0.02 \\ -0.01}$ \\
      \hline
      Electron spectral index & $2.13\substack{+0.03 \\ -0.04}$ & $2.38\substack{+0.02 \\ -0.02}$ \\
      \hline
      Proton spectral index  & $1.73\substack{+0.01 \\ -0.02}$ & X \\
      \hline
      electron energy cut-off (TeV) & $14.80\substack{+0.03 \\ -0.02}$ & $42.30\substack{+0.02 \\ -0.02}$ \\
      \hline
      proton energy cut-off (TeV) & $74.10\substack{+0.03 \\ -0.03}$ & X \\
      \hline
      B ($\mu$G) & $90.34\substack{+3.72 \\ -4.35}$ & $14.75\substack{+0.20 \\ -0.30}$ \\
      \hline
      Kep & $0.0075\substack{+0.0003 \\ -0.0004}$ & X \\
      \hline
      $n_H$ ($\text{cm}^{-3}$) & $39.89\substack{+0.83 \\ -0.96}$ & $0.07\substack{+0.04 \\ -0.04}$ \\
      \hline
      MaxLogLikelihood & $-37.69$ & $-46.32$ \\
      \hline
      \end{tabular}
    \caption{\label{tab:ParaFitRXJ}Parameters obtained from the fit for RX J1713.7-3946 using~Naima.}
\end{table}
\pagebreak

\begin{figure}
    \centering
    \begin{subfigure}{1\linewidth}
        \centering
        \includegraphics[width=0.6\linewidth, height = 0.4\linewidth]{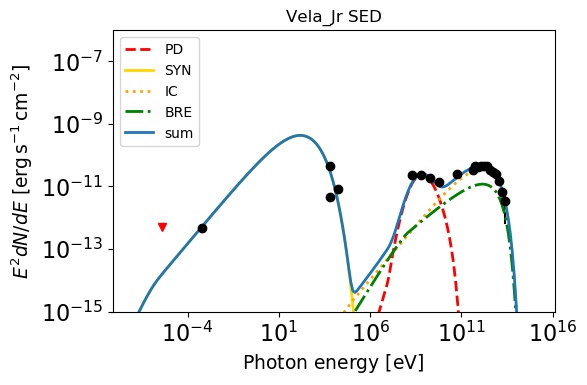}
        \caption{Vela Jr. fitted by using the lepto-hadronic~model.\label{VelaJ_H}}
    \end{subfigure}
    \begin{subfigure}{1\linewidth}
        \centering
        \includegraphics[width=0.6\linewidth, height = 0.4\linewidth]{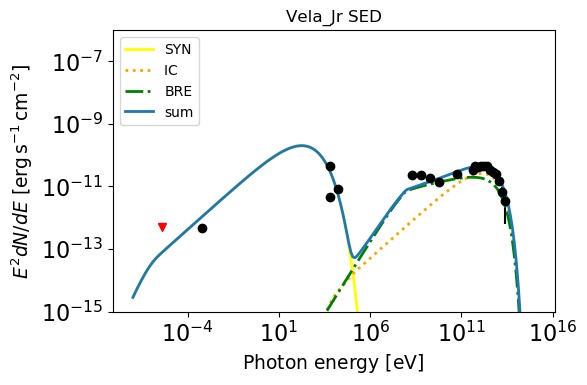}
        \caption{Vela Jr. fitted by using the leptonic~model.\label{VelaJ_L}}
    \end{subfigure}
    \caption{Naima fit results using different radiative scenarios for SNR Vela~Jr.}
\end{figure}

\begin{table}
    \centering
    \renewcommand{\arraystretch}{1.3}
    \begin{tabular}{ |p{6cm}|p{3cm}|p{3cm}|  }
      \hline
      Parameter & Lepto-hadronic & Leptonic \\
      \hline
      log10(Amplitude) & $46.70\substack{+0.02 \\ -0.02}$ & $44.80\substack{+0.02 \\ -0.02}$ \\
      \hline
      Electron spectral index & $1.61\substack{+0.03 \\ -0.03}$ & $1.96\substack{+0.02 \\ -0.02}$ \\
      \hline
      Proton spectral index  & $1.82\substack{+0.04 \\ -0.05}$ & X \\
      \hline
      electron energy cut-off (TeV) & $10.80\substack{+0.02 \\ -0.02}$ & $18.20\substack{+0.02 \\ -0.02}$ \\
      \hline
      proton energy cut-off (TeV) & $0.02\substack{+0.08 \\ -0.05}$ & X \\
      \hline
      B ($\mu$G) & $14.12\substack{+0.58 \\ -0.40}$ & $10.92\substack{+0.42 \\ -0.56}$ \\
      \hline
      Kep & $0.0112\substack{+0.0005 \\ -0.0003}$ & X \\
      \hline
      $n_H$ ($\text{cm}^{-3}$) & $46.28\substack{+6.46 \\ -2.74}$ & $84.72\substack{+9.59 \\ -10.57}$ \\
      \hline
      MaxLogLikelihood & $-2484.37$ & $-2536.26$ \\
      \hline
    \end{tabular}
    \caption{\label{tab:ParaFitVJ}Parameters obtained from the fit for Vela Jr. using~Naima.}
\end{table}
\pagebreak

\begin{figure}
    \centering
    \begin{subfigure}{1\linewidth}
        \centering
        \includegraphics[width=0.6\linewidth, height = 0.4\linewidth]{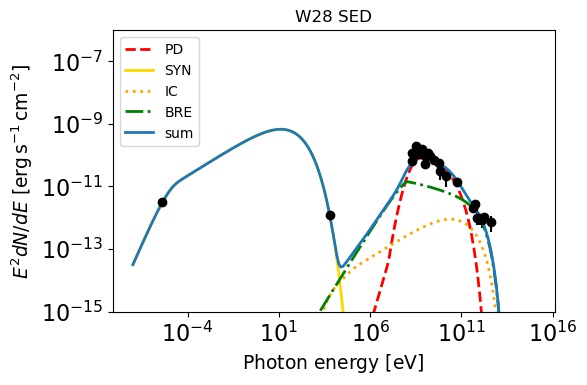}
        \caption{W 28 fitted by using the lepto-hadronic~model.\label{W28_H}}
    \end{subfigure}
    \begin{subfigure}{1\linewidth}
        \centering
        \includegraphics[width=0.6\linewidth, height = 0.4\linewidth]{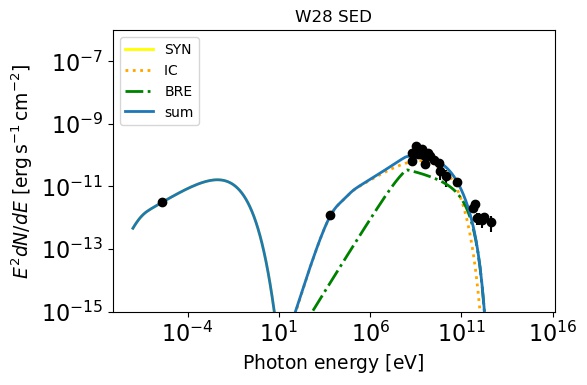}
        \caption{W 28 fitted by using the leptonic~model.\label{W28_L}}
    \end{subfigure}
    \caption{Naima fit results using different radiative scenarios for SNR W 28.}
\end{figure}

\begin{table}
    \centering
    \renewcommand{\arraystretch}{1.3}
    \begin{tabular}{ |p{6cm}|p{3cm}|p{3cm}|  }
      \hline
      Parameter & Lepto-hadronic & Leptonic \\
      \hline
      log10(Amplitude) & $47.81\substack{+0.04 \\ -0.04}$ & $48.63\substack{+0.10 \\ -0.10}$ \\
      \hline
      Electron spectral index & $2.27\substack{+0.02 \\ -0.03}$ & $2.29\substack{+0.04 \\ -0.03}$ \\
      \hline
      Proton spectral index  & $2.33\substack{+0.02 \\ -0.02}$ & X \\
      \hline
      electron energy cut-off (TeV) & $2.02\substack{+0.01 \\ -0.01}$ & $0.28\substack{+0.05 \\ -0.04}$ \\
      \hline
      proton energy cut-off (TeV) & $0.35\substack{+0.01 \\ -0.02}$ & X \\
      \hline
      B ($\mu$G) & $134.26\substack{+2.68 \\ -4.45}$ & $2.50\substack{+0.20 \\ -0.23}$  \\
      \hline
      Kep & $0.024\substack{+0.001 \\ -0.001}$ & X \\
      \hline
      $n_H$ ($\text{cm}^{-3}$) & $56.96\substack{+1.01 \\ -1.25}$ & $0.40\substack{+0.18 \\ -0.15}$ \\
      \hline
      MaxLogLikelihood & $-73.51$ & $-95.39$ \\
      \hline
    \end{tabular}
    \caption{\label{tab:ParaFitW28}Parameters obtained from the fit for W 28 using~Naima.}
\end{table}
\pagebreak

\begin{figure}
    \centering
    \begin{subfigure}{1\linewidth}
        \centering
        \includegraphics[width=0.6\linewidth, height = 0.4\linewidth]{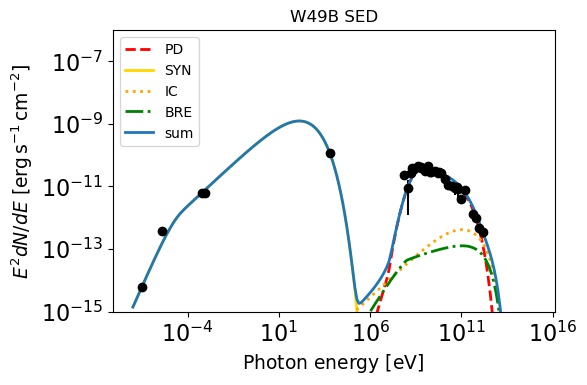}
        \caption{W 49B fitted by using the lepto-hadronic~model.\label{W49B_H}}
    \end{subfigure}
    \begin{subfigure}{1\linewidth}
        \centering
        \includegraphics[width=0.6\linewidth, height = 0.4\linewidth]{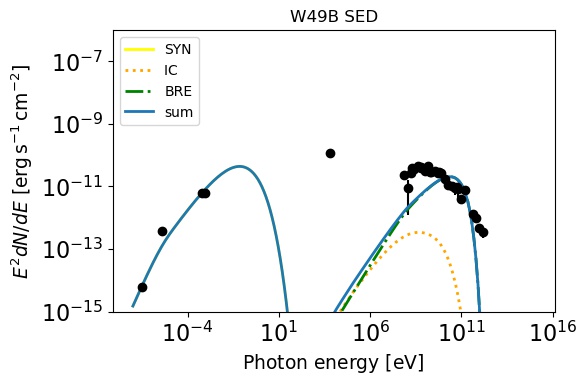}
        \caption{W 49B fitted by using the leptonic~model.\label{W49B_L}}
    \end{subfigure}
    \caption{Naima fit results using different radiative scenarios for SNR W 49B.}
\end{figure}

\begin{table}
    \centering
    \renewcommand{\arraystretch}{1.3}
    \begin{tabular}{ |p{6cm}|p{3cm}|p{3cm}|  }
      \hline
      parameters & Lepto-hadronic & Leptonic \\
      \hline
      log10(Amplitude) & $49.90\substack{+0.02 \\ -0.02}$ & $49.10\substack{+0.19 \\ -0.12}$ \\
      \hline
      Electron spectral index & $1.913\substack{+0.006 \\ -0.007}$ & $1.34\substack{+0.03 \\ -0.05}$ \\
      \hline
      Proton spectral index  & $2.29\substack{+0.02 \\ -0.01}$ & X \\
      \hline
      electron energy cut-off (TeV) & $3.10\substack{+0.01 \\ -0.02}$ & $0.10\substack{+0.03 \\ -0.05}$ \\
      \hline
      proton energy cut-off (TeV) & $1.60\substack{+0.003 \\ -0.005}$ & X \\
      \hline
      B ($\mu$G) & $252.87\substack{+4.91 \\ -3.29}$ & $55.04\substack{+6.54 \\ -7.91}$ \\
      \hline
      Kep & $0.0023\substack{+0.00004 \\ -0.0006}$ & X \\
      \hline
      $n_H$ ($\text{cm}^{-3}$) & $10.29\substack{+0.15 \\ -0.16}$ &  $205.86\substack{+32.48 \\ -18.30}$ \\
      \hline
      MaxLogLikelihood & $-128.37$ &$-570.01$ \\
      \hline
    \end{tabular}
    \caption{\label{tab:ParaFitW49B}Parameters obtained from the fit for W 49B using~Naima.}
\end{table}

\end{document}